\documentclass[final, 10pt]{article}
\pdfoutput=1
\usepackage{MRnotes}

\newcommand{\sVMS}{\widehat{\text{VMS}}}
\newcommand{\SF}[1]{\mathring{\mathscr{#1}}}
\newcommand{\Sum}[2]{\sum\limits_{#1}^{#2}}
\newcommand{\Prod}[2]{\prod\limits_{#1}^{#2}}

\newcommand{\zb}{\overline{z}}
\newcommand{\thetab}{\bar{\theta}}
\newcommand{\partialb}{\bar{\partial}}
\newcommand{\psit}{\widetilde{\psi}}
\newcommand{\chit}{\widetilde{\chi}}

\newcommand{\Lambdat}{\widetilde{\Lambda}}
\newcommand{\hP}{h_{_P}}
\newcommand{\hPt}{\widetilde{h}_{_P}}

\newcommand{\RsL}[2]{R^{#1}_{#2}}
\newcommand{\sNS}{S_{_\mathrm{NS}}}
\newcommand{\sR}{S_{_\mathrm{R}}}
\newcommand{\NNS}{N_{_{\mathrm{NS}}}}
\newcommand{\NR}{N_{_{\mathrm{R}}}}

\newcommand{\Ub}[1]{\Upsilon_b\pqty{#1}}
\newcommand{\UNS}[2]{\Upsilon^{(#1)}_{_\mathrm{NS}}\pqty{#2}}
\newcommand{\UNSo}{\Upsilon'_{_\mathrm{NS}}(0)}
\newcommand{\UR}[2]{\Upsilon^{(#1)}_{_\mathrm{R}}\pqty{#2}}
\newcommand{\URo}{\Upsilon'_{_\mathrm{R}}(0)}
\newcommand{\UNSh}[2]{\widehat{\mathsf{\Upsilon}}^{(#1)}_{_\mathrm{NS}}\pqty{#2}}
\newcommand{\URh}[2]{\widehat{\mathsf{\Upsilon}}^{(#1)}_{_\mathrm{R}}\pqty{#2}}
\newcommand{\dGb}[1]{\Gamma_b\pqty{#1}}
\newcommand{\dGNS}[2]{\Gamma^{(#1)}_{_\mathrm{NS}}\pqty{#2}}
\newcommand{\dGR}[2]{\Gamma^{(#1)}_{_\mathrm{R}}\pqty{#2}}

\newcommand{\CV}[1]{C^{(#1)}_{_\mathrm{V}}}
\newcommand{\CW}[1]{C^{(#1)}_{_\mathrm{W}}}
\newcommand{\Co}[1]{C^{(#1)}_\mathrm{odd}}
\newcommand{\Ce}[1]{C^{(#1)}_\mathrm{even}}
\newcommand{\CVdozz}{C_{_\mathrm{V,DOZZ}}}
\newcommand{\CWdozz}{C_{_\mathrm{W,DOZZ}}}
\newcommand{\Cedozz}{C^{\mathrm{even}}_{_\mathrm{DOZZ}}}
\newcommand{\Codozz}{C^{\mathrm{odd}}_{_\mathrm{DOZZ}}}

\newcommand{\drep}[2]{ \expval{ #1 , #2 } }
\newcommand{\PdegA}{P_{\drep{2}{1}}}
\newcommand{\PdegB}{P_{\drep{1}{2}}}

\newcommand{\rhoNS}{\rho_{_\mathrm{NS}}^{(b)}}
\newcommand{\rhoR}{\rho_{_\mathrm{R}}^{(b)}}

\newcommand{\bh}{\mathsf{b}}
\newcommand{\ch}{\mathsf{c}}
\newcommand{\Qh}{\mathsf{Q}}
\newcommand{\Ph}{\mathsf{P}}
\newcommand{\Vh}{\mathsf{V}}
\newcommand{\Wh}{\mathsf{W}}
\newcommand{\Lambdah}{\mathsf{\Lambda}}
\newcommand{\Lambdaht}{\widetilde{\mathsf\Lambda}}
\newcommand{\hh}{\mathsf{h}}
\newcommand{\hhP}{\mathsf{h}_{_\mathsf{P}}}
\newcommand{\hhPt}{\widetilde{\mathsf{h}}_{_\mathsf{P}}}
\newcommand{\RtL}[2]{\mathsf{R}^{#1}_{#2}}
\newcommand{\etaW}{\eta_{_\mathsf{W}}}
\newcommand{\etaR}{\eta_{\mathsf{R}}}

\newcommand{\CVh}[1]{\mathsf{C}^{(#1)}_{_\mathsf{V}}}
\newcommand{\CWh}[1]{\mathsf{C}^{(#1)}_{_\mathsf{W}}}
\newcommand{\Coh}[1]{\mathsf{C}^{(#1)}_\mathsf{odd}}
\newcommand{\Ceh}[1]{\mathsf{C}^{(#1)}_\mathsf{even}}

\title{Towards the super Virasoro minimal string }
\author{Mukund Rangamani${}^a$, Jianming Zheng${}^b$}

\affiliation[a]{
  Center for Quantum Mathematics and Physics (QMAP)\\
  Department of Physics \& Astronomy, University of California, Davis, CA 95616 USA}
\affiliation[b]{
  Department of Physics, Tsinghua University, Beijing 100084, China
}
\emailAdd{mukund@physics.ucdavis.edu}
\emailAdd{jmzphys@gmail.com}

\abstract{
The (bosonic) Virasoro minimal string, which relates worldsheet string theory to a deformation of the JT gravity matrix model, provides an interesting example of a tractable matrix/string 
duality. 
We explore its $\mathcal{N} =1$ supersymmetric generalization, the super Virasoro minimal string, which we expect to be dual to a deformation of the $\mathcal{N} =1$ JT supergravity matrix model.
The worldsheet theory is characterized by two copies of super Liouville theory, one with central charge $c > \frac{27}{2}$ (the spacelike regime) and another with $c < \frac{3}{2}$ (the timelike regime), coupled to worldsheet supergravity and subject to diagonal (Type 0A/B) GSO projection.  
As a first step, we define the timelike theory, which has hitherto not been bootstrapped, by obtaining its spectrum and structure constants.  Furthermore, we also outline the matrix model's predictions for the worldsheet observables. Curiously, all perturbative amplitudes are predicted to vanish in the 0B theory, while all tree-level amplitudes vanish in the 0A case.  Using the worldsheet description, we explicitly verify this prediction  (modulo an assumption) only for the simplest of the worldsheet observables, the sphere three-point function. A detailed study of other observables and verification of the duality is deferred for the future.}

\begin{document}
\maketitle


\section{Introduction}\label{sec:intro}

Non-critical string theories have provided invaluable insight into
non-perturbative dynamics. This is largely due to their being dual to (double-scaled) matrix models. These matrix string dualities are, in a certain sense, prototypical of holographic dualities such as AdS/CFT correspondence. As such, they have been instrumental in enabling us to decipher string dynamics beyond worldsheet CFT. For example, the celebrated $c=1$ string and its dual matrix models (cf.~\cite{Klebanov:1991qa,Ginsparg:1993is,Polchinski:1994mb} for reviews), provided insights into the strength of non-perturbative effects in the theory~\cite{Shenker:1990uf}. The developments in understanding quantum Liouville CFT~\cite{Dorn:1994xn,Zamolodchikov:1995aa} (see~\cite{Nakayama:2004vk} for a review) have enabled not only a detailed match of the perturbative observables~\cite{Balthazar:2017mxh}, but also engendered a deeper understanding of non-perturbative effects from the string theoretic description~\cite{Balthazar:2019rnh,Sen:2019qqg}.

In a related vein, the non-perturbative description of Jackiw-Teitelboim (JT) gravity as a random matrix model~\cite{Saad:2019lba} has provided insights into the  dynamics of low-dimensional quantum gravity. Moreover, it has been instrumental in clarifying various aspects of (near-extremal) black hole physics. 
For an overview of these developments, see the reviews~\cite{Mertens:2022irh,Turiaci:2024cad}.

An interesting one-parameter generalization of this duality is provided by the Virasoro minimal string (VMS)~\cite{Collier:2023cyw}. The worldsheet description involves a non-critical string background. The matter CFT consists of spacelike ($c > 25$) and timelike ($c < 1$) Liouville CFTs. Both of these are coupled to worldsheet gravity (the bosonic $\mathfrak{b}\mathfrak{c}$-ghost CFT). The results of~\cite{Collier:2023cyw} form a rich tapestry linking worldsheet strings, matrix models, 3d gravity, and intersection theory on moduli spaces.
Specifically, the worldsheet description is dual to a double-scaled matrix integral, whose eigenvalue density, $\rho_{_\mathrm{VMS}}(E) = \frac{2\,\sqrt{2}}{\sqrt{E}}\, \sinh(2\pi\,b\,\sqrt{E})\, \sinh(2\pi\,b^{-1}\,\sqrt{E})$, is a one-parameter deformation (the parameter being $b$) of the JT eigenvalue density $\rho_{_\mathrm{JT}}(E) =\sinh(2\pi\, \sqrt{E})$, the latter being obtained from the former in the semiclassical limit $b\to 0$.

In the current paper, we lay the groundwork for generalizing the VMS to
include worldsheet supersymmetry. Specifically, we will be interested in
coupling $\mathcal{N} = 1$ super Liouville theory with
$c > \frac{27}{2}$ (the \emph{spacelike super Liouville} theory) and an analogous version with $c < \frac{3}{2}$ (the \emph{timelike super Liouville} theory) to two-dimensional $\mathcal{N}  = (1,1)$ supergravity on the worldsheet. This worldsheet string admits a  single  fermion number operator, $(-1)^{F_\mathrm{ws}}$, but not individual holomorphic and anti-holomorphic fermion number operators. Therefore, one can only implement a non-chiral worldsheet GSO projection and thereby construct the Type 0A/B strings from these building blocks. We will refer to this construction as the $\mathcal{N}=1$ super Virasoro minimal string ($\sVMS$). One might naively expect such worldsheet strings to be related to deformations of  matrix models dual to JT supergravity. This class of matrix models were analyzed in detail in~\cite{Stanford:2019vob}, the salient features of which we will review below (some  aspects of non-perturbative physics were discussed~\cite{Johnson:2021owr}).

The Type 0 strings,  we recall, retain just the NS-NS and R-R states and project out the R-NS sector, thereby having no fermions in the target space spectrum. This can be justified by constructing a modular invariant partition function by the diagonal sum over spin structures~\cite{Seiberg:1986by}. For strings propagating in the critical dimension (with say flat spacetime target $\mathbb{R}^{9,1}$), one retains the (NS$+$, NS$+$), (NS$-$, NS$-$) sector including the tachyon from the latter (see~\cite{Polchinski:1998rr} for an overview). The distinction between the 0A and 0B theory lies in what R-R states one keeps. In the latter one retains the (R$+$, R$+$) and (R$-$, R$-$) sectors, while in the former we retain (R$+$, R$-$) and (R$-$, R$+$). A similar statement applies in the context of the non-critical $\hat{c} =1$ string theory\footnote{We reserve the hat decoration to denote the rescaled central charge of superconformal theories, and refrain from using it to denote observables and parameters of the  timelike super Liouville theory, cf.~\cref{fn:sstL}.}~\cite{Takayanagi:2003sm,Douglas:2003up}.  These theories involve coupling $\mathcal{N} =1$ super Liouville theory with $c=\frac{27}{2}$ ($\hat{c} =9$) with a single free boson (the target spacetime coordinate). The Type 0B theory in this context is dual to a large $N$ limit of a $\mathrm{U}(N)$ gauged matrix quantum mechanics. The Type 0A theory, on the other hand, is related to a quiver matrix quantum mechanics theory with gauge group $\mathrm{U}(N) \times \mathrm{U}(N+q)$, with $q$ parameterizing the background flux.
Recently, the perturbative and non-perturbative dynamics of the 
Type 0B string theory were examined 
in~\cite{Balthazar:2022atu,Balthazar:2022apu} (we will find their results helpful in our analysis).

The worldsheet string we seek to construct is closer in spirit to the minimal
strings, where one couples Liouville theory to a minimal model. In the
supersymmetric case, such minimal string theories were analyzed
in~\cite{Klebanov:2003wg,Seiberg:2004at}. Following the connection between
minimal bosonic strings and JT gravity developed in~\cite{Seiberg:2019upl,Mertens:2020hbs}, it was argued in~\cite{Mertens:2020pfe} that a similar story ought to hold for minimal superstrings. Specifically, the relation between super-JT and minimal string disk amplitudes was investigated and found to match. Another piece of evidence comes from~\cite{Fredenhagen:2007tk} which argues that 
the $c\to \frac{3}{2}$ limit super Virasoro minimal models can be identified with a continuation of super Liouville theory. Our goal here is to develop the necessary technology to analyze the worldsheet super Virasoro minimal string, and lay the foundations for a detailed duality with a matrix model description.

Let us first take stock of the ingredients involved in the worldsheet
construction. The primary ingredient for us is the $\mathcal{N} =1$ spacelike super Liouville theory. The classical theory was constructed in~\cite{Distler:1989nt}, but we will be interested in the quantum description as a two-dimensional SCFT. For this purpose, it will suffice to specify the operator spectrum, which like in bosonic Liouville theory is continuous, and the structure constants. The presence of fermionic degrees of freedom implies that we have to keep track of both Neveu-Schwarz (NS) and Ramond (R) sectors.
Fortunately, this theory has been extensively analyzed in the literature for the spacelike regime $c > \frac{27}{2}$. In
fact, following the solution of the bosonic Liouville theory by DOZZ~\cite{Dorn:1994xn,Zamolodchikov:1995aa}, the structure constants of the super Liouville theory were initially bootstrapped in~\cite{Rashkov:1996np,Poghossian:1996agj}. They were subsequently reanalyzed in~\cite{Fukuda:2002bv} who also examined the boundary states. We shall review salient aspects of these works below (cf.~\cref{sec:quantumSL}).

The second ingredient we require is a theory with $c < \frac{3}{2}$ to combine with the spacelike super Liouville theory and make up the net central charge of $15$ for the
superstring. While, as mentioned earlier, one could follow the logic of the minimal string construction, and pick a unitary supersymmetric minimal model, we will instead want to work with a super Liouville theory with central charge $c < \frac{3}{2}$.  The semiclassical theory was recently analyzed in~\cite{Anninos:2023exn}, but our interest is once again in the quantum SCFT. This regime will be referred to as the timelike super Liouville theory.

We will specify the quantum theory by simply writing down a spectrum and structure constants that satisfy crossing symmetry. The logic is similar to that encountered in bosonic timelike Liouville theory. The crossing symmetry constraints from correlators of degenerate operators leads to a functional equation for the structure constants. In addition to the original DOZZ solution, there is a second solution to these bootstrap
conditions~\cite{Schomerus:2003vv,Zamolodchikov:2005fy,Kostov:2005kk}. Loosely speaking, this defines a certain analytic continuation of Liouville theory, which has now been understood
from several perspectives~\cite{Harlow:2011ny,Ribault:2015sxa,Bautista:2019jau}. 
By specifying the spectrum and structure constants, and ensuring that the latter satisfy crossing, one defines the quantum theory. The timelike Liouville theory, a bosonic non-unitary CFT, thus characterized, was one of the principal ingredients of the VMS. 

We likewise seek a second solution for crossing constraints arising from degenerate 4-point correlators for super Liouville theory. Specifically, we demonstrate the existence of a second solution for the structure constants satisfying the functional relations, which are derived by adapting Teschner's trick~\cite{Teschner:1995yf} to the supersymmetric case. This is largely facilitated by the earlier analysis of~\cite{Poghossian:1996agj,Fukuda:2002bv},
who examined degenerate correlators in super Liouville theory. 
The determination of the structure constants for the timelike theory constitutes one of our primary results.  
We also undertake a numerical check of the crossing equations for physical correlators (in some subsectors) and demonstrate consistency. As an upshot, we define the timelike super Liouville SCFT by specifying the operator content and the structure constants. These results can be viewed as a continuation from real values of $b$, which parameterizes the spacelike theory, to purely imaginary values  $b \to -  i\,b$ to define the timelike case. For clarity, we parameterize the timelike theory by an a priori independent parameter $\bh$.\footnote{We  adapt a convention where the parameters (and operators) defining the timelike Liouville theory are denoted with a sans serif font, viz., by $\{\bh,\Qh,\ch,\Ph\}$ etc.\label{fn:sstL}} The timelike and spacelike parameters will be related when we consider the worldsheet string construction. 

To check crossing in the supersymmetric theory, one needs information about the $\mathcal{N}=1$ superconformal blocks. This can be done using recursion in the central charge or internal operator weight along the lines described in~\cite{Zamolodchikov:1987avt} for bosonic conformal blocks. The essential results were derived in~\cite{Hadasz:2006qb, Belavin:2007gz,Hadasz:2007nt} for the Neveu-Schwarz sector and in~\cite{Hadasz:2008dt,Suchanek:2010kq} for the Ramond sector.  
The thesis~\cite{Suchanek:2009ths} provides a nice compilation of the salient results.  In addition, a  numerical check of crossing equations for the spacelike regime was
carried out in~\cite{Suchanek:2010kq}. 
Recently, these structure constants have been exploited in analyzing the perturbative S-matrix of the $c=1$ string~\cite{Balthazar:2022atu}. They also numerically analyzed the crossing symmetry for the NS sector external states using the aforementioned results for the superconformal blocks.  Adapting their analysis for the timelike case,\footnote{Crossing symmetry for timelike bosonic Liouville theory (and also for complex values of $b$) was numerically checked in the bosonic case initially in~\cite{Ribault:2015sxa}.} we verify our
prediction for the structure constants. Specifically, we have ascertained crossing to hold for the NS sector structure constants for a range of external weights.

Having a definition of the (quantum) timelike super Liouville theory, we construct the worldsheet Type 0 $\sVMS$ string theory. As we explain, while certain aspects of the analysis are a straightforward generalization of the VMS, we will encounter some subtleties in formulating a complete story. To explain the issues encountered, we first note that since we are only checking crossing for sphere correlators, our structure constants are determined only up to an overall sign. This will be important comparing of the string construction to matrix models.

To gain some insight, let us turn to the matrix models dual to JT supergravity. As indicated at the outset, these (and other generalizations) were analyzed thoroughly in~\cite{Stanford:2019vob}.  Our interest is in the 0A and 0B supergravity theories, which we expect to obtain from our string construction in the semiclassical ($b\to 0$) limit. These two theories are distinguished by the nature of the spin structure sum. The distinction lies in whether a non-anomalous $(-1)^F$ symmetry survives after the sum over spin structures.\footnote{ A succinct summary of the salient features of supermoduli space volumes with $\mathcal{N} = 1$ can be found in the review~\cite{Turiaci:2024cad}.}

In the 0B theory, there is no $(-1)^F$ symmetry. The Hamiltonian in this case is the square of the (self-adjoint) supercharge. This supercharge is drawn from a GUE ensemble. The spectral curve is given by $\rho_{_\mathrm{0B}} = \frac{1}{\pi}\,\cosh (2\pi\,q)$ with $E= q^2$. Due to the absence of any edge to the spectral curve, all the perturbative observables, which one expects to compute the volumes of the $\mathcal{N}=1$ moduli space, vanish.  In the 0A case, the $(-1)^F$ operator survives, and correspondingly, the dual matrix model is drawn from the $(1,2)$ ensemble. In this case, one can get non-vanishing answers. This is because the theory by definition includes a factor of $(-1)^\zeta$, where $\zeta$ is the  mod 2 index.\footnote{We thank Douglas Stanford and Edward Witten for clarifying remarks on this point. } While the sphere amplitudes still vanish owing to the presence of Grassmann odd moduli, higher genus volumes are non-vanishing.

The spectral curve in the case of the $\sVMS$ should be a one-parameter
deformation. A natural guess is that it is given by the universal density of
states of the $\mathcal{N}=1$ superconformal theory~\cite{Collier:2023cyw}. This would suggest a
spectral curve $\rho \sim \frac{1}{\sqrt{E}}\, \cosh(\pi\,b\,\sqrt{E})\, \cosh (\pi\,b^{-1}\, \sqrt{E})$. Again, such a matrix model would predict vanishing perturbative observables. Certain aspects of this matrix model were analyzed in~\cite{Johnson:2024fkm} where the vanishing of perturbative observables was noted and certain non-perturbative effects analyzed. However, this work solely focuses on the matrix model taking the aforementioned guess for the spectral curve at face value. Importantly, belying the title,  does not examine the  worldsheet construction of the supersymmetric Virasoro minimal strings. 

Taking this matrix model results into account, we encounter a somewhat curious situation. The Type 0B string has all of its perturbative amplitudes vanishing, while the Type 0A string has all tree level amplitudes vanishing. This structure ought to be reproduced from the worldsheet computation. Unfortunately, we are unable to offer firm evidence that this is the case at present, and will defer a detail analysis for the future. Specifically, we encounter an issue already at the level of the three-point sphere amplitude, which should be related  to the volume of the
three-holed sphere. The vanishing in the supergravity limit follows from the presence of Grassmann odd moduli. However, in the worldsheet correlator, we find that the result depends on some sign choices for the timelike super Liouville structure constant. We fix our signs by demanding consistency with the matrix model prediction, and provide a heuristic argument for this choice.  Note that even if we could provide a rationale for this tree-level amplitude to vanish, our choice of sign doesn't explain why higher point amplitudes (and higher-loop amplitudes) vanish. Be that as it may, for a worldsheet theory to only have non-perturbative observables that are non-vanishing suggests a deeper principle that we appear to be missing. In light of this, we leave a detailed analysis of the worldsheet observables to the future.

The outline of the paper is as follows. We begin in~\cref{sec:spL} with a brief overview of the spacelike $\mathcal{N}=1$ Liouville theory. We not only summarize the spectrum and structure constants, but also provide a short synopsis of the derivation of the latter. In~\cref{sec:tsL} we then turn to the timelike case, and demonstrate that there is a second solution to crossing that satisfies various properties. We outline the numerical checks we have done to verify our prediction for the structure constants. With the two SCFTs at hand, we briefly describe in~\cref{sec:svms} the worldsheet construction, and explain how the genus-0 three-point function can be seen to be consistent with the matrix model prediction. We conclude in~\cref{sec:discuss} with an overview of open issues. 

The appendices contain some technical results: ~\cref{sec:splfns} collates properties of the Barnes double-Gamma and Upsilon functions, while ~\cref{sec:norms} compares the normalization of structure constants we define with those in the literature. In~\cref{sec:sward} we compile superconformal Ward identities for the three-point functions. Finally, in~\cref{sec:superrecurion} we outline the essential features of the
$\mathcal{N}=1$ superconformal blocks. 

\medskip
\noindent \emph{Note added:} During the course of our analysis we became aware that Beatrix M\"uhlmann, Vladimir Narovlansky, and Ioannis Tsiares have independently derived the structure constants for the timelike super Liouville theory and are investigating the worldsheet description of the super Virasoro minimal string. 
Their results are being released concurrently with ours~\cite{Muhlmann:2025tdb}. We are  
grateful to them for reaching out and informing us of their analysis, generously sharing their insights, and for coordinating the submission of both sets of results.

\section{The \texorpdfstring{$\mathcal{N}=1$}{N=1} spacelike
  super Liouville theory}\label{sec:spL}

We will begin with an overview of the spacelike $\mathcal{N} =1$
super Liouville theory, which, as we noted
in~\cref{sec:intro}, is fairly well understood. Our
goal here is to outline the essential points succinctly. This will be helpful first in
our analysis of the timelike theory, and thence in the worldsheet string construction.

We will work in $\mathcal{N} = (1,1)$ superspace with coordinates $z, \zb, \theta , \thetab$, and let $D = \partial_\theta + \theta\, \partial_z$ be the homomorphic super derivative (similarly, $\overline{D} = \partial_{\thetab} + \thetab\, \partialb_{\zb}$). The basic Liouville superfield is
\begin{equation}\label{eq:Lsf}
\mathring{\Phi}  = \phi + i\, \theta\, \psi + i\, \thetab\, \psit + i\, \theta\,\thetab\, F_\mathrm{aux}\,.
\end{equation}
The classical action for the super Liouville theory is given by~\cite{Distler:1989nt}
\begin{equation}\label{eq:SLsfaction}
S = \frac{1}{2\pi} \int d^2 z\, d^2 \theta \left( D \mathring{\Phi}\, \overline{D} \mathring{\Phi} + 4\pi i\, \mu_s\, e^{b\,\mathring{\Phi}} \right),
\end{equation}
where $\mu_s$ is the cosmological constant, and $b$ is the Liouville parameter.
Integrating over the Grassmann  coordinates, and eliminating the auxiliary field $F_\mathrm{aux}$,
we arrive at the standard classical action (in flat space)
\begin{equation}\label{eq:sLaction}
S = \frac{1}{2\pi} \int\, d^2z\, \pqty{
  \partial \phi\, \partialb\phi +
  \psit \,\partial \psit + \psi \,\partialb \psi +
  4\pi i\,\mu_s\,b^2\, \psit\,\psi\, e^{b\phi} + 4\pi^2\, \mu_s^2\, b^2\,
  e^{2\,b\,\phi}  } \,.
\end{equation}
The classical stress tensor and supercurrent (holomorphic components) are given by
\begin{equation}
\begin{split}
T & =
-\frac{1}{2}\, \pqty{\partial \phi\, \partial \phi - Q\, \partial^2\, \phi
+ \psi\, \partial \psi} \\
T_F & =
i\, \pqty{\psi\, \partial \phi - Q\, \partial \psi}\,.
\end{split}
\end{equation}
%

\subsection{The quantum theory: spectrum and structure constants}\label{sec:quantumSL}

The quantum dynamics of the aforementioned action is an $\mathcal{N} =1$ SCFT
with $c \geq \frac{27}{2}$.
As in the bosonic case, the parameter $b$ determines the background Liouville
charge $Q$ and the central charge $c$ through
\begin{equation}\label{eq:QbcsL}
Q = b + \frac{1}{b} \,,\qquad
c = \frac{3}{2}\, \hat{c} =  \frac{3}{2} + 3\, Q^2\,.
\end{equation}

The $\mathcal{N} =1$ Liouville SCFT is characterized by its primary operators,
which lie in the Neveu-Schwarz (NS) and Ramond (R) sectors depending on the
fermion boundary conditions. The spectrum of these primaries is continuous,
parameterized by a Liouville momentum $P \in \mathbb{R}_+$. Specifying the
operator spectrum and the structure constants suffices to characterize the
theory. The results presented below were originally obtained
in~\cite{Poghossian:1996agj,Rashkov:1996np,Fukuda:2002bv}, and a
useful summary can be found in~\cite{Balthazar:2022atu}.

\paragraph{The Neveu-Schwarz sector:} The operator content in this sector is built atop a  superconformal primary $V_P$, which in the semiclassical limit can be identified as $e^{\pqty{\frac{Q}{2} \pm i\,P}\, \phi}$.
The supermultiplet containing this operator can be characterized as 
\begin{equation}\label{eq:Snsop}
\SF{S}_P = V_P +  \theta\, \Lambda_P +  \thetab\, \Lambdat_P - \theta\,\thetab \, W_P \,.
\end{equation}
The conformal weights of these operators are\footnote{
  A note on conventions: when multiple momenta (indexed by $P_i$) are involved, we will alleviate notation by writing $h_i \equiv h_{P_i}$. Similarly, it will sometimes be useful to introduce
  $\alpha_i \equiv \frac{Q}{2} + i\,P_i$ to keep expressions
  readable. \label{fn:notation} }
\begin{equation}
(\hP ,\hPt), 
\quad 
\pqty{\hP +\frac{1}{2}, \hPt}, 
\quad
\pqty{\hP, \hPt + \frac{1}{2}},  
\quad  
\pqty{\hP +\frac{1}{2}, \hPt + \frac{1}{2}}\,,
\end{equation}
respectively, with
\begin{equation}\label{eq:sLhp}
\hP= \hPt = \frac{1}{2}\, \pqty{\frac{Q^2}{4} + P^2} \,.
\end{equation}

\paragraph{The Ramond sector:}
In the Ramond sector, the operators are constructed from the spin $\sigma$ and disorder operators $\mu$ of the free fermion theory. Each has weight $(\frac{1}{16}, \frac{1}{16})$. The $\sigma$ field is $\mathbb{Z}_2$ odd and $\mu$ is the operator at the end of the $\mathbb{Z}_2$ topological defect line. These are dressed with the Liouville field and lead to two R sector operators $\RsL{\pm}{P}$. The signs indicate the eigenvalue of the fermion number  operator of the spacelike theory, which we denote as
$(-1)^{F_s}$. In the asymptotic regime we can identify these operators as 
\begin{equation}\label{eq:Srop}
\RsL{+}{P} \sim \sigma\, e^{\pqty{\frac{Q}{2} \pm i\,P }\, \phi}
\,,\qquad
\RsL{-}{P} \sim  \mu\, e^{\pqty{\frac{Q}{2} \pm i\,P }\, \phi} \,.
\end{equation}
They both have weights 
\begin{equation}
\pqty{\hP + \frac{1}{16}, \hPt + \frac{1}{16}}
= 
\pqty{\frac{c}{24} + P^2, \frac{c}{24} +P^2}\,.
\end{equation}
We note in passing that~\cite{Fukuda:2002bv} derives the structure constants by working with chiral twist fields $s^\pm$ and $\tilde{s}^\pm$, defining the left-right combinations  
\begin{equation}\label{eq:chiraltwist}
\begin{split}
\Theta^{\pm\pm}_P 
&\sim 
    e^{\pm i\frac{\pi}{4}}\, s^\pm\, \tilde{s}^\pm \, e^{(\frac{Q}{2}+iP)\phi}\,, \\
\Theta^{\pm\mp}_P
&\sim 
    s^\pm\, \tilde{s}^\mp \, e^{(\frac{Q}{2}+iP)\phi} \,.
\end{split}
\end{equation}
This choice can be mapped to the fields we use through the identification  
\begin{equation}\label{eq:chiralsymmap}
\RsL{+}{P} = \frac{e^{-i\,\frac{\pi}{4}}}{\sqrt{2}}\, \pqty{\Theta^{++}_P  
    +\Theta^{--}_P}  \,, \qquad  
\RsL{-}{P}  = \frac{1}{\sqrt{2}}\,\pqty{\Theta^{+-}_P + \Theta^{-+}_P } \,. 
\end{equation}

\paragraph{The normalization of the two-point functions:}
To characterize the SCFT we need to
specify the structure constants. Before doing so let us first normalize the
two-point functions of the operators as follows
\begin{equation}\label{eq:ns2ptnorm}
\expval{V_{P_1}(0)\, V_{P_2}(1)} = \frac{1}{\rhoNS(P_1)}\,
\pqty{\delta(P_1  - P_2) + \delta(P_1 + P_2)} \,.
\end{equation}
Here $\rhoNS$ is the NS sector spectral density obtained from the modular
crossing kernel of the $\mathcal{N}=1$ vacuum character in the NS sector, $\Tr_{_\mathrm{NS}}(q^{L_0})$ and is given by~\cite{Mertens:2017mtv}
\begin{equation}\label{eq:rhons}
\rhoNS(P) = 4\, \sinh(\pi\,b\, P)\, \sinh(\pi\,b^{-1}\, P)\,.
\end{equation}
The normalization of operators in the NS supermultiplet is then inherited
from this by the action of the supercurrent, e.g.,
\begin{equation}
\begin{split}
\expval{W_{P_1}(0)\, W_{P_2}(1)}
 & = -
4\, h_1^2 \,
\expval{V_{P_1}(0)\, V_{P_2}(1)}\,.
\end{split}
\end{equation}

In the Ramond sector we have instead
\begin{equation}\label{eq:r2ptnorm}
\expval{\RsL{\pm}{P_1}(0)\, \RsL{\pm}{P_2}(1)} = \frac{1}{2\,\rhoR(P_1)}
\pqty{\delta(P_1 - P_2) \pm \delta(P_1 + P_2)}\,.
\end{equation}
Now $\rhoR$ is the R sector spectral density obtained from the modular
crossing kernel of the $\mathcal{N}=1$ vacuum character in the NS sector with periodic fermion boundary conditions, i.e.,
$\Tr_{_\mathrm{NS}} ((-1)^F\, q^{L_0})$ ~\cite{Mertens:2017mtv}
and is given by 
\begin{equation}\label{eq:rhor}
\rhoR(P) = 2\sqrt{2}\, \cosh(\pi\,b\, P)\, \cosh(\pi\,b^{-1}\, P)\,.
\end{equation}

\paragraph{The structure constants:} There are four independent three-point functions that characterize the theory. All of them involve at least one insertion of the superconformal primary $V_P$. Furthermore, in three of the structure constants, one of the operators is distinct from the other two, whose argument we will distinguish when necessary.  As in the bosonic Liouville theory, they are given in terms of the Barnes double-gamma function $\dGb{z}$. It is useful to define the combinations 
\begin{equation}\label{eq:dgNSR}
\begin{split}
\dGNS{b}{z}
 & \equiv
\dGb{\frac{z}{2}}\, \dGb{\frac{z+Q}{2}} \,, \\
\dGR{b}{z}
 & \equiv
\dGb{\frac{z+b}{2}}\, \dGb{\frac{z+b^{-1}}{2}}\,.
\end{split}
\end{equation}
Some basic properties of $\dGb{z}$ are compiled in~\cref{sec:splfns} for quick reference.  For a more comprehensive summary of the building blocks, we refer the reader to the appendices of~\cite{Eberhardt:2023mrq}.

Two of the structure constants involve the spinless NS sector operators and are given to be
\begin{equation}\label{eq:CVWdef}
\begin{split}
\expval{V_{P_1}(0)\, V_{P_2}(1)\, V_{P_3}(\infty)}
 & =
\CV{b}(P_1,P_2,P_3)\,, \\
\expval{W_{P_1}(0)\, V_{P_2}(1)\, V_{P_3}(\infty)}
 & =
\CW{b}(P_1, P_2,P_3)\,.
\end{split}
\end{equation}
We don't distinguish the location of $W_P$ in $\CW{b}$ as the result will turn out to be symmetric in the three arguments.  
Choosing a reflection symmetric normalization\footnote{ The structure constants in the conventional (i.e., DOZZ) normalization are collated in~\cref{sec:norms}.} we have
\begin{equation}\label{eq:sLCV}
\begin{split}
\CV{b}(P_1,P_2,P_3)
 & = \frac{\dGNS{b}{2\,b +2\,b^{-1}}}{2\, \dGNS{b}{b+b^{-1}}^3}\,
\frac{\Prod{\epsilon_{1,2,3}{} = \pm 1}
\, \dGNS{b}{\frac{b+b^{-1}}{2} + i\,  \Sum{j=1}{3}\, \epsilon_j\, P_j}}{
\Prod{k=1}{3}\, \Prod{\epsilon_k= \pm 1}{}\,
\dGNS{b}{b+b^{-1} + 2\,i\, \epsilon_k\,P_k}} \,,                  \\
 & =
\frac{\dGNS{b}{2\,Q}}{2\, \dGNS{b}{Q}^3}\,
\frac{\dGNS{b}{\frac{Q}{2} \pm i\,P_1 \pm i\,P_2 \pm i\,P_3}}{\Prod{k=1}{3}\,
  \dGNS{b}{Q \pm 2\,i\, P_k}} \,.
\end{split}
\end{equation}
In the first equality we make it explicit that there are eight terms
in the numerator, and a pair of terms for each external momentum
label $k$ in the denominator. The second line writes this out in a
commonly used shorthand form, where we take the product over all the permutations of the signs involved. Furthermore, the r.h.s.\ should be viewed as a function of $b$ and $P_i$. In particular, we should replace all occurrences of $Q$ in terms of $b$ using~\eqref{eq:QbcsL}, as is again indicated in the first line. This will become relevant when we present our results for the timelike case. In what follows, we will use the shorthand notation for brevity, with the above expression serving to remind us of its meaning.

With these conventions in place, the second structure constant in the NS sector is given by 
\begin{equation}\label{eq:sLCW}
\CW{b}(P_1,P_2,P_3) = i\,\frac{\dGNS{b}{2\,Q}}{\dGNS{b}{Q}^3}\,
\frac{\dGR{b}{\frac{Q}{2} \pm i\,P_1 \pm i\,P_2 \pm i\,P_3}}{\Prod{k=1}{3}\,
  \dGNS{b}{Q \pm 2\,i\, P_k}}\,.
\end{equation}
As noted earlier, this structure constant is symmetric in the three momenta. 
Correlation functions involving other operators in the NS
supermultiplet~\eqref{eq:Snsop} can be obtained using superconformal Ward identities, cf.~\cref{sec:sward}. 
In particular, all the NS sector 3-point correlators are fixed terms of $\CV{b}$ and $\CW{b}$.

Turning to the Ramond sector, structure constants involving an odd number of R operators vanish. The non-vanishing three-point functions involve mixed NS and R correlators,
and can be determined to be the following
\begin{equation}\label{eq:Ceodef}
\begin{split}
\expval{V_{P_1}(0)\, \RsL{+}{P_2}(1)\, \RsL{+}{P_3}(\infty)}
 & = \frac{1}{2} \,
\pqty{\Ce{b}(P_1;P_2,P_3) + \Co{b}(P_1;P_2,P_3) } \,, \\
\expval{V_{P_1}(0)\, \RsL{-}{P_2}(1)\, \RsL{-}{P_3}(\infty)}
 & = \frac{1}{2} \,
\pqty{\Ce{b}(P_1;P_2,P_3) - \Co{b}(P_1;P_2,P_3) }\,.
\end{split}
\end{equation}
Since the position of the NS operator $V_P$ in the correlator is distinguished, we singled out its momentum label in expression for the structure constants, separating it with a semicolon from the other two as indicated. 
The functions $\Ce{b}$ and $\Co{b}$ are themselves given as
\begin{equation}\label{eq:sLCeo}
\begin{split}
\Ce{b}(P_1;P_2,P_3)
 & =
\frac{\dGNS{b}{2\,Q}}{2\, \dGNS{b}{Q}^3}\,
\frac{\dGR{b}{\frac{Q}{2} \pm i\, (P_1+P_2+P_3)}\, \dGR{b}{\frac{Q}{2} \pm
i\, (P_1 - P_2-P_3)}}{\dGNS{b}{Q \pm 2\,i\, P_1}} \\
 & \qquad \qquad \times
\frac{\dGNS{b}{\frac{Q}{2} \pm i\, (P_1-P_2 + P_3)}\,
  \dGNS{b}{\frac{Q}{2} \pm i\,(P_1 + P_2 - P_3}}{\dGR{b}{Q\pm 2\,i\, P_2}\,
\dGR{b}{Q\pm 2\,i\, P_3}}\,,                      \\
\Co{b}(P_1;P_2,P_3)
 & =
\frac{\dGNS{b}{2\,Q}}{2\, \dGNS{b}{Q}^3}\,\frac{\dGNS{b}{\frac{Q}{2} \pm i\, (P_1+P_2+P_3)}\, \dGNS{b}{\frac{Q}{2} \pm
i\, (P_1 - P_2-P_3)}}{\dGNS{b}{Q \pm 2\,i\, P_1}} \\
 & \qquad \qquad  \times
\frac{\dGR{b}{\frac{Q}{2} \pm i\, (P_1-P_2 + P_3)}\,
  \dGR{b}{\frac{Q}{2} \pm i\,(P_1 + P_2 - P_3}}{\dGR{b}{Q\pm 2\,i\, P_2}\,
  \dGR{b}{Q \pm2\,i\, P_3}}\,.
\end{split}
\end{equation}

In presenting the structure constants, we have chosen a normalization
convention analogous to the one employed in~\cite{Collier:2023cyw} for bosonic
Liouville theory. The relation between our choice and the conventional
presentation is explained in~\cref{sec:norms}. Specifically, taking $P_1 \to i\, \frac{Q}{2}$ we find
\begin{equation}\label{eq:sLC1limit}
\begin{split}
\lim_{P_1 \to i\,\frac{Q}{2}}\, \CV{b}(P_1,P_2,P_3)
 & = \frac{\delta(P_2 - P_3)}{\rhoNS(P_2)} \,, \\
\lim_{P_1 \to i\,\frac{Q}{2}}\, \Ce{b}(P_1;P_2,P_3)
 & = \frac{\delta(P_2 - P_3)}{\rhoR(P_2)}\,,   \\
\lim_{P_1 \to i\,\frac{Q}{2}}\, \Co{b}(P_1;P_2,P_3)
 & = \frac{\delta(P_2 + P_3)}{\rhoR(P_2)} \,.
\end{split}
\end{equation}
This verifies that we recover the two-point functions with our chosen normalization as specified in~\eqref{eq:ns2ptnorm} and~\eqref{eq:r2ptnorm}, respectively. On the other hand, $\CW{b}$ vanishes if we analytically continue one of the $V$ operators to the identity, since there is no two-point function between $V_P$ and $W_P$.

\subsection{A sketch of the derivation of the structure constants}\label{sec:sL3pt}

To obtain the structure constants quoted above, the  basic idea is to
consider the 4-point function involving one degenerate operator and derive a recursion relation as in~\cite{Teschner:1995yf}. For the supersymmetric theory, this was analyzed in~\cite{Poghossian:1996agj,Rashkov:1996np,Fukuda:2002bv}. We briefly review the essential elements in what follows.

The degenerate operators of $\mathcal{N} =1$ superconformal algebra exist
at specific values of the Liouville momentum. These occur at
\begin{equation}
P_{\drep{r}{s}} =  \frac{i}{2}\, \pqty{r\, b +  \frac{s}{b}}\,, \qquad
r,s \in \mathbb{Z}_{>0}\,.
\end{equation}
Alternately, using the parameterization $\alpha = \frac{Q}{2} + i\,P$ (cf.~\cref{fn:notation}) the null states are parameterized as
$\alpha_{\drep{r}{s}}=\frac{1}{2}\, (Q-r\,b-s\,b^{-1})$.
The null states are at level $\frac{1}{2}\,r\,s$. For $r +s$ odd, the states belong to the R sector, while for $r+s$ even, they lie in the NS sector. The first two non-trivial degenerate operators, with
momenta $\PdegA$ and $\PdegB$, are the R-sector operators
$\RsL{\pm}{\PdegA}$ and $\RsL{\pm}{\PdegB}$, respectively. They  are annihilated by the following linear combination of the  (homomorphic) super Virasoro generators
\begin{equation}
L_{-1}-\frac{2\,b^2}{2\,b^2+1}G_{-1}\,G_0 \,, \qquad
L_{-1} - \frac{2}{2 + b^2}\, G_{-1}\, G_0 \,.
\end{equation}

The recursion relations are obtained by considering the degenerate 4-point functions with an insertion of say
$\RsL{\delta}{\PdegA}$, with $\delta \in \{\pm1\}$, As an explicit example, consider the correlator 
\begin{equation}\label{eq:VVRR}
\mathfrak{C}_\mathrm{4,deg}(z)
= \expval{V_{P_3}(z_3)\, V_{P_2}(z_2)\, \RsL{\delta}{\PdegA}(w)\, \RsL{\delta}{P_1}(z_1)}\,,
\end{equation}
where $z$ is the cross-ratio of the variables $\{z_3,z_2,w,z_3\}$.
One uses the fusion of the degenerate operator $\RsL{\delta}{\PdegA}(w)$ with $V_P(z_2)$ or with $\RsL{\delta}{P_1}(z_1)$. There are finitely many terms in this degenerate OPE, which is determined to be   
\begin{equation}
\begin{split}
\RsL{\delta}{\PdegA}(w)\, V_P(0)
 & \sim
\abs{w}^{b\,(\frac{Q}{2} +i\,P) }\, 
\RsL{\delta}{P+i\,\frac{b}{2}}(0)
+\delta\, F_{_\mathrm{RNS}}(P)\, 
\abs{w}^{b\,(\frac{Q}{2} -i\,P) }\, 
\RsL{\delta}{P -i\,\frac{b}{2}}(0) \,, \\
\RsL{\delta}{\PdegA}(w)\, \RsL{\delta}{P}(0)
 & \sim
\abs{w}^{b\,(\frac{Q}{2} +i\,P)+\frac{3}{4} }\, 
W_{P+i\,\frac{b}{2}}(0) 
+ \frac{i}{4}\, \abs{w}^{b\,(\frac{Q}{2} +i\,P) - \frac{1}{4} }\, V_{P +i\,\frac{b}{2}}(0)
  \\
& 
+ F_{_\mathrm{RR}}(P)\, \bqty{
\abs{w}^{b\,(\frac{Q}{2} -i\,P) - \frac{1}{4} }\, 
V_{P -i\,\frac{b}{2}}(0)
+\frac{\abs{w}^{b\,(\frac{Q}{2} -i\,P) + \frac{3}{4}}}{4\,(Q-b - 2\,i\,P)^2} 
\, 
W_{P -i\,\frac{b}{2}}(0) } \,.
\end{split}
\end{equation}
The coefficients in the fusion can be obtained by exploiting the free field limit using the Coulomb gas
formalism. They are 
\begin{equation}\label{eq:RVRfusion}
\begin{split}
F_{_\mathrm{RNS}}(P)
 & = \mu\,\pi \,b^2\, \gamma\pqty{\frac{bQ}{2}}\,
\gamma\pqty{\frac{1-b^2}{2} -i\,b\,P}\, \gamma\pqty{i\,b\,P} \,, \\
F_{_\mathrm{RR}}(P)
 & =
 2\,i\, F_{_\mathrm{RNS}}(-P + \frac{i\,b}{2}) \,,
\end{split}
\end{equation}
where $\gamma(z)$ is defined as a ratio of Gamma functions, 
viz.,
\begin{equation}\label{eq:gammadef}
\gamma(z)  = \frac{\Gamma(z)}{\Gamma(1-z)}\,.
\end{equation}
The fusion of $\RsL{\delta}{\PdegA}$ with $\RsL{\delta'}{P}$  with $\delta \neq \delta'$ involve the level half descendants of $V_P$, viz., $\Lambda_P$ and $\Lambdat_P$ and can be determined using superconformal transformations. 

The idea is to use the null vector decoupling equation to write down the general solution to~\eqref{eq:VVRR}. One exploits the fusion rules and the fact that the solution can be as a combination of hypergeometric functions (which are the degenerate conformal blocks). From here, one derives a set of functional equations for suitable ratios of the structure constants, which are then solved to obtain the analog of the DOZZ formula.  We spell out the essential steps without dwelling on the details (which can be found in ~\cite{Poghossian:1996agj,Rashkov:1996np,Fukuda:2002bv}) for completeness. 

For instance, for the NS structure constants, we use the 
$\RsL{\delta}{\PdegA}\, \RsL{\delta'}{P}$ fusion to arrive at a relation between the structure constants $\CW{b}$ and $\CV{b}$.
In this case, dropping overall factors, one can write the result for the degenerate correlator (swapping $P$ for $\alpha$ as indicated in~\cref{fn:notation} for brevity) as  
\begin{equation}
\begin{split}
\mathfrak{C}_\mathrm{4,deg}(z)
&\propto  
    -\frac{\CW{b}(P_1 +i\, \frac{b}{2}, P_2, P_3)}{\pqty{\frac{Q}{2} +i\,P_1-\,\frac{b}{2}}^2}\,  
    \abs{G_{1}(\alpha_1,\alpha_2, \alpha_3; z)}^2 \\
&\qquad    
    +  F_{_\mathrm{RR}}(\alpha_1)\,\CV{b}(P_1 -i\, \frac{b}{2}, P_2, P_3)\, 
    \abs{G_{2}(Q-\alpha_1,\alpha_2, \alpha_3; z)}^2 \,.
\end{split}
\end{equation}
The functions $G_{1,2}$ are standard hypergeometric functions, 
\begin{equation}
\begin{split}
G_{1}(\alpha_1,\alpha_2, \alpha_3; z)
&= 
    z^{\frac{b\,\alpha_1}{2} + \frac{3}{8}}\, (z)^{\frac{b\,\alpha_2}{2}}\, 
    {}_2F_1\pqty{\frac{i\,b\,P_{1+2+3}}{2} + \frac{3}{4}
    ,\frac{i\,b\,P_{1+2-3}}{2} + \frac{3}{4}, i\,b\,P_1 + \frac{3}{2}; z } \,,
\\
G_{2}(\alpha_1,\alpha_2, \alpha_3; z)
&= 
    z^{\frac{b\,\alpha_1}{2} - \frac{3}{8}}\, (1-z)^{\frac{b\,\alpha_2}{2}}\, 
    {}_2F_1\pqty{\frac{i\,b\,P_{1+2+3}}{2} + \frac{1}{4}
    ,\frac{i\,b\,P_{1+2-3}}{2} + \frac{1}{4}, i\,b\,P_1 + \frac{1}{2}; z } \,.
\end{split}
\end{equation}
We employ a shorthand notation $P_{i+j\pm k} = P_i + P_j \pm P_k$ for convenience. 

One can now exploit the crossing formulae for the hypergeometric functions relating ${}_2F_1\pqty{a,b,c;z}$ and 
${}_2F_1\pqty{a',b',b',1-z}$ to obtain the correlator in the crossed channel. Finally, requiring that the resulting crossed correlator is single valued gives a functional relation between the two structure constants appearing in the degenerate correlator. 

We will write this recursion relation after stripping off an  overall normalization factor for each of the vertex operators. This allows us to determine the non-trivial momentum independent part of the structure constants. One may then fix the normalizations by demanding that they too satisfy the recursion, and ensuring further the correct limiting behavior~\eqref{eq:sLC1limit} is attained. Therefore, in the formulae below, we indicate ratios of structure constants with an explicit subscript `\st{norm}' to emphasize that  that normalizations are factored out. 

For the NS structure constants, one finds relations of the form  
\begin{equation}\label{eq:CWCVratios}
\begin{split}
\mathfrak{N}_\epsilon(P_1,P_2,P_3) 
&=\equiv
\frac{\CW{b}(P_1 + \epsilon\, \frac{i\,b}{2}, P_2, P_3)}{\CV{b}(P_1- \epsilon\, \frac{i\,b}{2},P_2, P_3)}\Bigg|_{\text{\st{norm}}} \\ 
&=
b^{-2-4i\,\epsilon\,b\,P_1}\, \prod_{\epsilon_{2,3} = \pm 1}  \gamma\pqty{\frac{3}{4} + \frac{i\, b}{2}\, \pqty{\epsilon\,P_1 + \epsilon_2 \, P_2 + \epsilon_3\, P_3}}\,.
\end{split}
\end{equation}
The prefactors have been judiciously chosen to require compatibility with the recursion of the Upsilon functions. The dual relation with $b \to 1/b$ is obtained by using the degenerate operator $\RsL{\pm}{\PdegB}$.

All told, we have four functional equations satisfied by the functions $\CV{b}$ and $\CW{b}$. The solution
presented in~\eqref{eq:sLCV} and~\eqref{eq:sLCW} exploits the functional relations satisfied by the Upsilon functions~\eqref{eq:dGfrel}. One engineers a combination of these functions and picks a suitable normalization to arrive at the final result quoted earlier.

The analysis for the structure constants with Ramond operators is similar, and involves using the $\RsL{\delta}{\PdegA}\, V_P$ fusion to derive functional relations for the ratio of $\Ce{b}$ and $\Co{b}$. The relations are 
\begin{equation}\label{eq:CeCoratios}
\begin{split}
\mathfrak{R}_+(P_1,P_2,P_3)
&\equiv  
\frac{\Ce{b}(P_1 ; P_2+ \frac{i\,b}{2}, P_3)}{\Co{b}(P_1, P_2 -\frac{i\,b}{2},P_3)}\Bigg|_{\text{\st{norm}}} \\ 
&=
b^{-4i\,b\,P_2} 
\prod_{\epsilon_{1} = \pm 1}  \gamma\pqty{\frac{3}{4} + \frac{i\, b}{2}\, \pqty{\epsilon_1\,P_1 +  P_2 +  P_3}}\,
\gamma\pqty{\frac{1}{4} + \frac{i\, b}{2}\, \pqty{\epsilon_1\,P_1 +  P_2 - P_3}}\,, \\
\mathfrak{R}_-(P_1,P_2,P_3)
&\equiv 
\frac{\Co{b}(P_1 ;P_2+\frac{i\,b}{2},P_3)}{\Ce{b}(P_1 ; P_2 -\frac{i\,b}{2},P_3)}\Bigg|_{\text{\st{norm}}} \\
&=
b^{-4i\,b\,P_2} 
\prod_{\epsilon_{1} = \pm 1}  \gamma\pqty{\frac{3}{4} + \frac{i\, b}{2}\, \pqty{\epsilon_1\,P_1 +  P_2 -  P_3}}\,
\gamma\pqty{\frac{1}{4} + \frac{i\, b}{2}\, \pqty{\epsilon_1\,P_1 +  P_2 + P_3}}\,.
\end{split}
\end{equation}
To derive them, we found it convenient to work with the chiral twist fields used in~\cite{Fukuda:2002bv} as described in~\eqref{eq:chiraltwist}, and thence re-express the result for the Ramond vertex operators. 

\medskip
\begin{figure}[ht]
\begin{subfigure}[a]{0.5\linewidth}
\centering
    \includegraphics[width=0.9\textwidth]{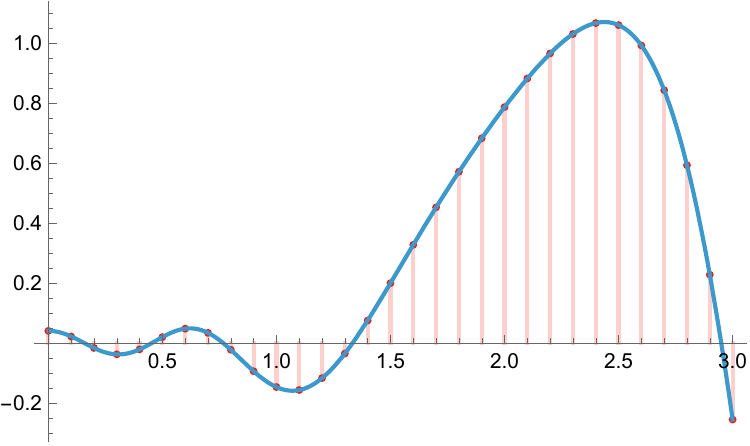}
\begin{picture}(0.3,0.4)(0,0)
    \put(5,10){\makebox(0,0){$\scriptstyle{P_1}$}}
    \put(-100,100){\makebox(0,0){$\scriptstyle{\Re \mathfrak{N}_+(P_1)}$}}
\end{picture}
\end{subfigure}
\begin{subfigure}[a]{0.5\linewidth}
    \includegraphics[width=0.9\textwidth]{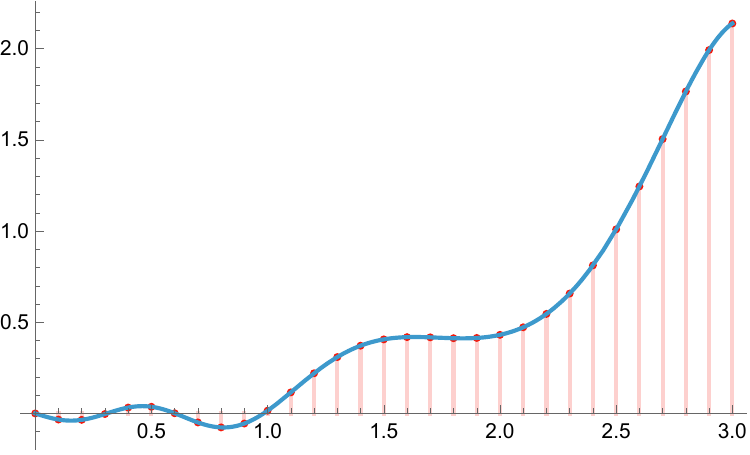}
\begin{picture}(0.3,0.4)(0,0)
    \put(5,10){\makebox(0,0){$\scriptstyle{P_1}$}}
    \put(-100,100){\makebox(0,0){$\scriptstyle{\Im \mathfrak{N}_+(P_1)}$}}
\end{picture}
\end{subfigure}
\bigskip
\vspace{2cm}
\begin{subfigure}[a]{0.5\linewidth}
    \includegraphics[width=0.9\textwidth]{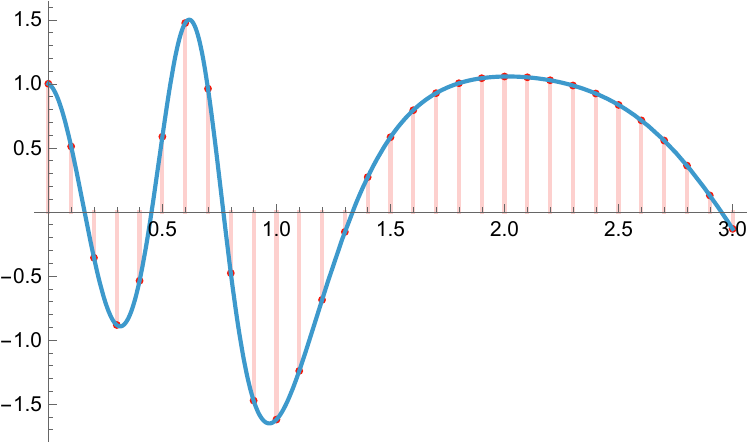}
 \begin{picture}(0.3,0.4)(0,0)
    \put(0,50){\makebox(0,0){$\scriptstyle{P_2}$}}
    \put(-100,100){\makebox(0,0){$\scriptstyle{\Re \mathfrak{R}_+(P_2)}$}}
\end{picture}
\end{subfigure}
\begin{subfigure}[a]{0.5\linewidth}
    \includegraphics[width=0.95\textwidth]{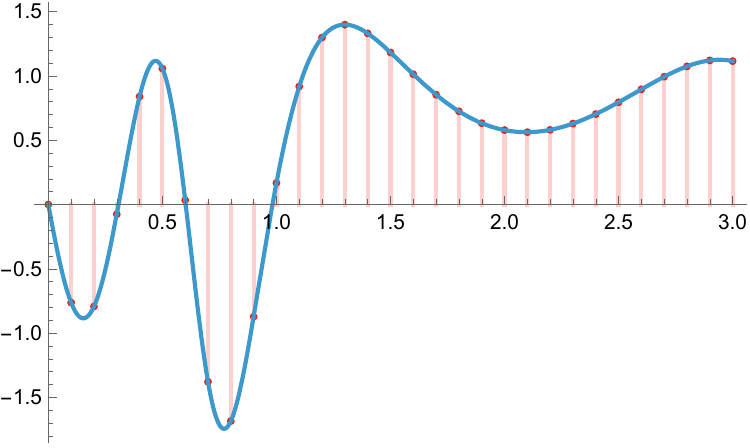}
\begin{picture}(0.3,0.4)(0,0)
    \put(0,50){\makebox(0,0){$\scriptstyle{P_2}$}}
    \put(-80,110){\makebox(0,0){$\scriptstyle{\Im \mathfrak{R}_+(P_2)}$}}
\end{picture}
\end{subfigure}
\vspace{-2cm}
\caption{A simple numerical verification of the functional relations for the ratios of structure constants for $b = \frac{\pi}{2}$, i.e., for $c=16.1181$. The top row shows the real and imaginary parts for the ratio of NS sector correlators $\mathfrak{N}_+(P_1, \frac{1}{2}, \frac{1}{6})$.
The bottom row depicts the ratio of R sector correlators $\mathfrak{R}_+(\frac{1}{2},P_2,\frac{1}{6})$. In each case, the solid curve is the product of $\gamma$-functions and the discrete data points are obtained by plotting the product of double-Gamma functions appearing in the structure constants. Plots for $\mathfrak{N}_-$ and $\mathfrak{R}_-$ are similar and therefore not depicted. }
\label{fig:slrecursioncheck}
\end{figure}
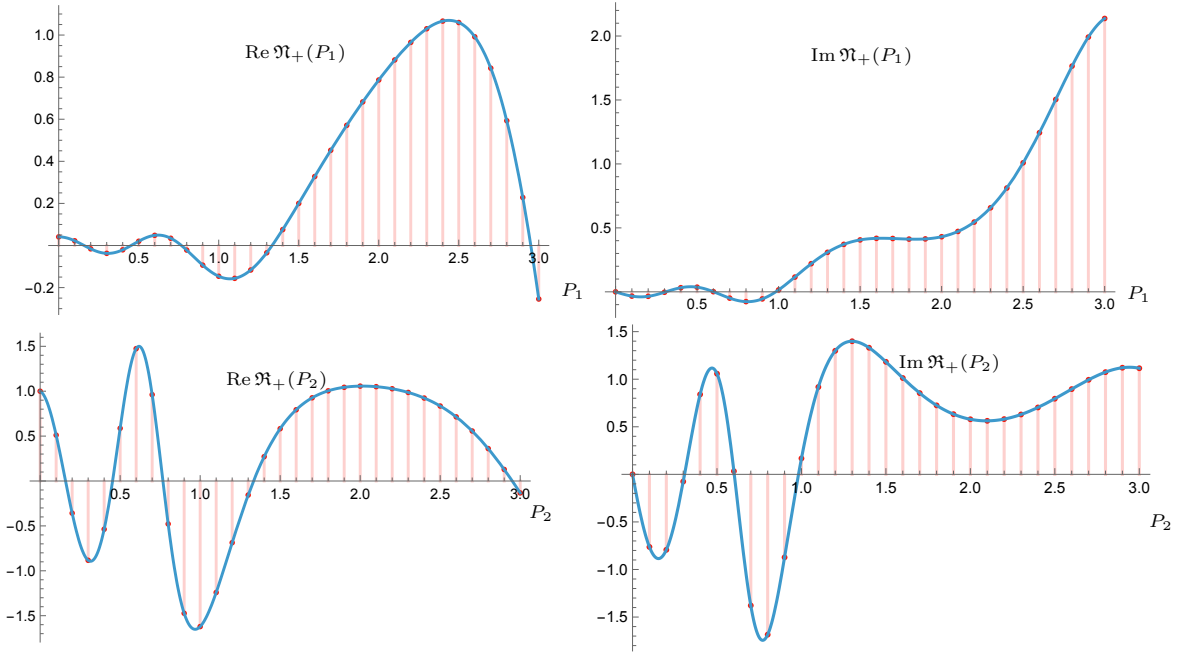

One can verify using the relations~\eqref{eq:dGfrel} satisfied by $\dGb{z}$ that the product of the functions in the numerator of~\eqref{eq:sLCV},~\eqref{eq:sLCW},~\eqref{eq:sLCeo} that depend on all three momenta satisfy these constraints. We also have checked this numerically (mostly to calibrate our analysis in the timelike case) and illustrate an example check in~\cref{fig:slrecursioncheck}.
To evaluate the function $\dGb{z}$ for generic values of $b$ we used the approximations described in~\cite{Ribault:2015sxa}.

\section{The \texorpdfstring{$\mathcal{N}=1$}{N=1} timelike super Liouville theory}\label{sec:tsL}

We would now like to turn to the timelike super Liouville theory, which we would like to define for $c \leq \frac{3}{2}$. The  fields characterizing this theory can be obtained from~\eqref{eq:Lsf} by analytically continuing the fields of the spacelike Liouville theory, $(\phi, \psi,\psit, F_\mathrm{aux}) \to i\,(\xi, \chi,\chit,F_\mathrm{aux}')$, which we package into a superfield
\begin{equation}\label{eq:tLsuperfield}
\mathring{\Xi} = \xi + i\,\theta\, \chi + i\,\thetab\, \chit + i\, \theta\, \thetab\, 
F_\mathrm{aux}' \,.
\end{equation}
Classically, one may write the following action
\begin{equation}\label{eq:tLaction}
S = \frac{1}{2\pi} \int\, d^2z\, \pqty{
  - \partial \xi\, \partialb\xi -
  \chit \,\partial \chit - \chi \,\partialb \chi +
  4\pi i\,\mu_t\,\bh^2\, \chit\,\chi\, e^{\bh\,\xi} + 4\pi^2\, \mu_t^2\,
  \bh^2\, e^{2\,\bh\,\xi}  } \,.
\end{equation}
In particular, we have also analytically continued the Liouville parameter $b \to -i\,\bh$. A useful reference for aspects of the classical theory and some semiclassical physics is~\cite{Anninos:2023exn}.

\subsection{The quantum timelike super Liouville theory}\label{sec:tsLquantum}

The quantum theory is defined with by a parameter $\bh\in \mathbb{R}_+$, with
\begin{equation}
\Qh = \bh^{-1}- \bh\,, \qquad \ch = \frac{3}{2}- 3\,\Qh^2 \ \leq \ \frac{3}{2}\,.
\end{equation}
We will find it convenient to write formulae directly in terms of
the special functions used in the spacelike case. Since the latter
was defined in terms of $(b,Q)$, we will often invoke the following spacelike to timelike analytic continuation $b\to -i\, \bh$ and
$Q \to i\, \Qh$. However, the structure constants, as we shall see below, are not analytically continued from the spacelike case (as in the bosonic Liouville theory).

\paragraph{The spectrum of operators:} 
The spectrum comprises operators in the NS and R-R sectors as before, indexed by a Liouville momentum $\Ph$. The NS supermultiplet is
\begin{equation}\label{eq:Tnsop}
\SF{T}_{\Ph} = \Vh_{\Ph} +  \theta\, \Lambdah_{\Ph} +  \thetab\, \Lambdaht_{\Ph} - \theta\,\thetab \, \Wh_{\Ph} \,.
\end{equation}
The conformal weights of these operators are%
\begin{equation}
(\hhP ,\hhPt), \quad \pqty{\hhP +\frac{1}{2}, \hhPt}, \quad
\pqty{\hhP, \hhPt + \frac{1}{2}},  \quad  \pqty{\hhP +\frac{1}{2}, \hhPt + \frac{1}{2}}\,,
\end{equation}
respectively, with ($\Ph \in \mathbb{R}_+$)
\begin{equation}\label{eq:tLhp}
\hhP= \hhPt = \frac{1}{2}\, \pqty{-\frac{\Qh^2}{4} + \Ph^2} \,.
\end{equation}

In the Ramond sector,  we once again dress the spin $\sigma$ and disorder operators $\mu$ of the free fermion theory with the timelike Liouville field, resulting
\begin{equation}\label{eq:TLrop}
\RtL{+}{\Ph} \sim \sigma\, e^{\pqty{\frac{\Qh}{2}\pm i\,\Ph }\, \xi}
\,,\qquad
\RtL{-}{\Ph} \sim  \mu \, e^{\pqty{\frac{\Qh}{2}\pm i\,\Ph }\, \xi} \,,
\end{equation}
both of which have weights,
\begin{equation}
\pqty{\hhP + \frac{1}{16}, \hhPt + \frac{1}{16}}\,.
\end{equation}
Note that we are not distinguishing the spin and disorder operators from the ones used in the spacelike theory. When we combine the two sets of super Liouville theories, we will disambiguate them.

\paragraph{The structure constants:}
Having laid out the spectrum, we need to specify the structure constants.
The superconformal Ward identities continue to apply and reduce the number of independent structure constants to four. These are the analogs of~\eqref{eq:CVWdef} and~\eqref{eq:Ceodef} which we characterize in the timelike case as  
\begin{equation}\label{eq:CtLVeoWdef}
\begin{split}
\expval{\Vh_{\Ph_1}(0)\, \Vh_{\Ph_2}(1)\, \Vh_{\Ph_3}(\infty)}
 & =
\CVh{\bh}(\Ph_1,\Ph_2,\Ph_3)\,, \\
\expval{\Wh_{\Ph_1}(0)\, \Vh_{\Ph_2}(1)\, \Vh_{\Ph_3}(\infty)}
 & =
\CWh{\bh}(\Ph_1,\Ph_2,\Ph_3)\,,\\
\expval{\Vh_{\Ph_1}(0)\, \RtL{+}{\Ph_2}(1)\, \RtL{+}{\Ph_3}(\infty)}
 & = \frac{1}{2} \,
\pqty{\Ceh{\bh}(\Ph_1;\Ph_2,\Ph_3) + \Coh{\bh}(\Ph_1;\Ph_2,\Ph_3) } \,, \\
\expval{\Vh_{\Ph_1}(0)\, \RtL{-}{\Ph_2}(1)\, \RtL{-}{\Ph_3}(\infty)}
 & = \frac{1}{2} \,
\pqty{\Ceh{\bh}(\Ph_1;\Ph_2,\Ph_3) - \Coh{\bh}(\Ph_1;\Ph_2,\Ph_3) }\,.
\end{split}
\end{equation}

Once again, these structure constants should be fixed by demanding that they satisfy the superconformal bootstrap constraints. We claim the solution arising from imposing the said constraints to be the following:
\begin{equation}\label{eq:tLCs}
\begin{split}
\CVh{\bh}(\Ph_1,\Ph_2,\Ph_3)
 & = \frac{2i}{\CW{\bh}(i\,\Ph_1,i\,\Ph_2, i\,\Ph_3)}  \,, \\
\CWh{\bh}(\Ph_1,\Ph_2,\Ph_3)
 & =  \frac{2i\,\etaW}{\CV{\bh}(i\,\Ph_1,i\,\Ph_2, i\,\Ph_3)} \,,\\
\Ceh{\bh}(\Ph_1,\Ph_2,\Ph_3)
 & = \frac{1}{\Co{\bh}(i\,\Ph_1,i\,\Ph_2, i\,\Ph_3)}  \,,  \\
\Coh{\bh}(\Ph_1,\Ph_2,\Ph_3)
 & = \frac{\etaR}{\Ce{\bh}(i\,\Ph_1,i\,\Ph_2, i\,\Ph_3)}
\end{split}
\end{equation}
In writing these expressions, we have explicitly acknowledged the fact that some overall (momentum independent) signs are left undetermined by the functional relations. These are indicated by the coefficients $\etaW$ and $\etaR$.  We will later furnish argue for the choice $\etaW  = -1$, and 
$\etaR = 1$, respectively.

To present the result in a compact form, we have expressed our result in terms of the functions appearing in the spacelike theory. The latter are to be viewed as functions of the Liouville momenta, and the Liouville parameter $b$. In particular, one should replace $Q \to Q(b)$ in these expressions before rewriting them in terms of the timelike Liouville parameter $\bh$. To be clear,  it is worth recording one of these expressions to make our notation transparent. For example, the NS structure constants are explicitly given by 
\begin{equation}\label{eq:CVWtLexplicit}
\begin{split}
\CVh{\bh}(\Ph_1,\Ph_2,\Ph_3)
&= 
\frac{2\, \dGNS{\bh}{\bh+\bh^{-1}}^3}{\dGNS{\bh}{2\,\bh+2\,\bh^{-1}}}\,
\frac{\Prod{k=1}{3}\,
\dGNS{\bh}{\bh+\bh^{-1} \pm 2\,\Ph_k}}{\dGR{\bh}{\frac{\bh+\bh^{-1}}{2} \pm \Ph_1 \pm \Ph_2 \pm \Ph_3}} \,,\\
\CWh{\bh}(\Ph_1,\Ph_2,\Ph_3)
&= 4i\,\etaW \, \frac{\dGNS{\bh}{\bh+\bh^{-1}}^3}{\dGNS{\bh}{2\,\bh+2\,\bh^{-1}}}\,
\frac{\Prod{k=1}{3}\,
\dGNS{\bh}{\bh+\bh^{-1} \pm 2\,\Ph_k}}{\dGNS{\bh}{\frac{\bh+\bh^{-1}}{2} \pm \Ph_1 \pm \Ph_2 \pm \Ph_3}} \,.
\end{split}
\end{equation}
The determination of the structure constants for the timelike theory~\eqref{eq:tLCs} is one of our primary results.

Notice that, as in the bosonic case, the timelike structure constants are not analytic continuations of the spacelike ones. Not only, are the momenta continued to imaginary values, but there is also a swap of the two combinations of the Upsilon functions tagged by the NS and R labels.  In addition, when one computes higher-point correlators, we need to specify the contour of integration for the momenta $\Ph_i$. These are taken to lie on a shifted real axis with a small imaginary part $\Ph \in \mathbb{R}_+ + i\,\varepsilon_\Ph$ to avoid the poles along the real axis present in the structure constants. This discussion is largely similar to the one in the bosonic case~\cite{Ribault:2015sxa}, and the essential point is well explained in~\cite{Collier:2023cyw}. 

\subsection{Checks of the timelike structure constants}\label{sec:tsLchecks}

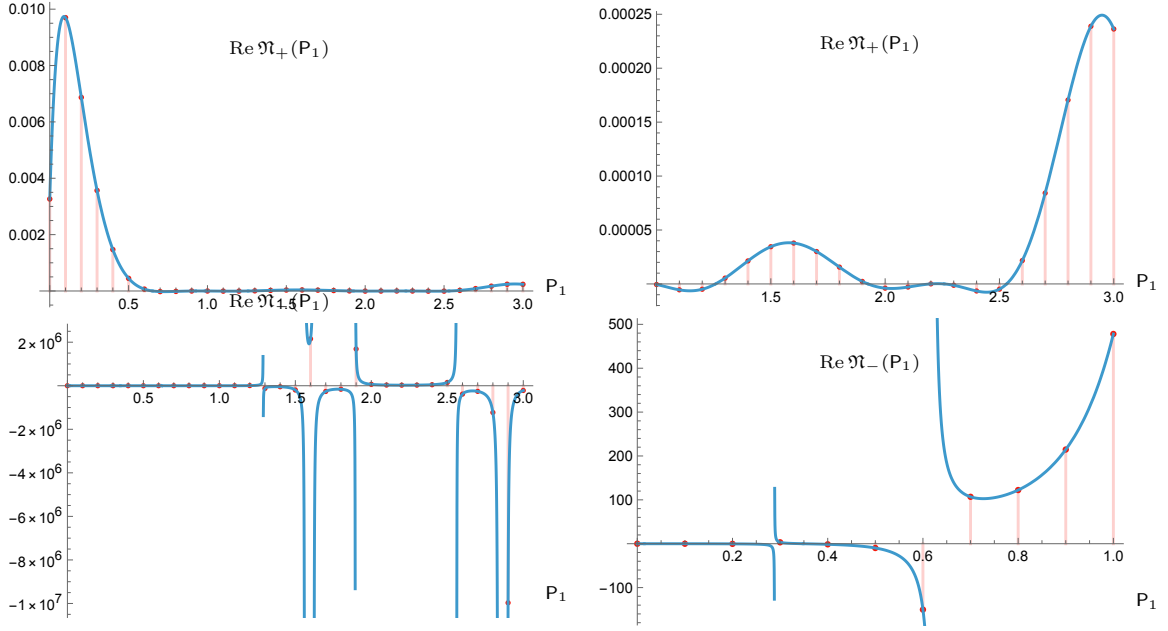
\begin{figure}[htp!]
\begin{subfigure}[a]{0.5\linewidth}
\centering
    \includegraphics[width=0.9\textwidth]{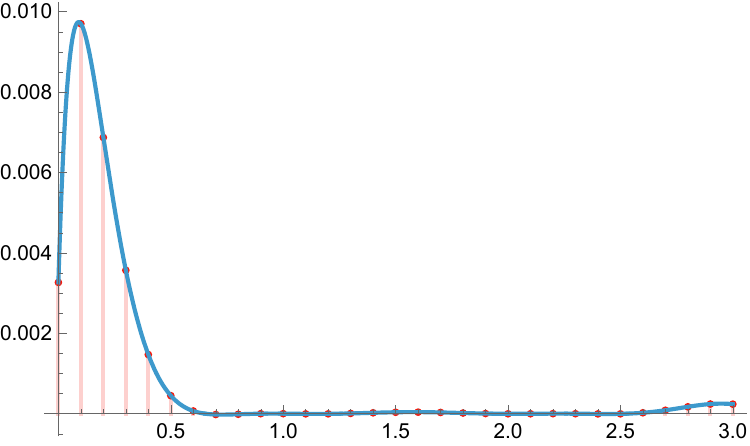}
\begin{picture}(0.3,0.4)(0,0)
    \put(5,10){\makebox(0,0){$\scriptstyle{\Ph_1}$}}
    \put(-100,100){\makebox(0,0){$\scriptstyle{\Re \mathfrak{N}_+(\Ph_1)}$}}
\end{picture}
\end{subfigure}
\begin{subfigure}[a]{0.5\linewidth}
\centering
    \includegraphics[width=0.9\textwidth]{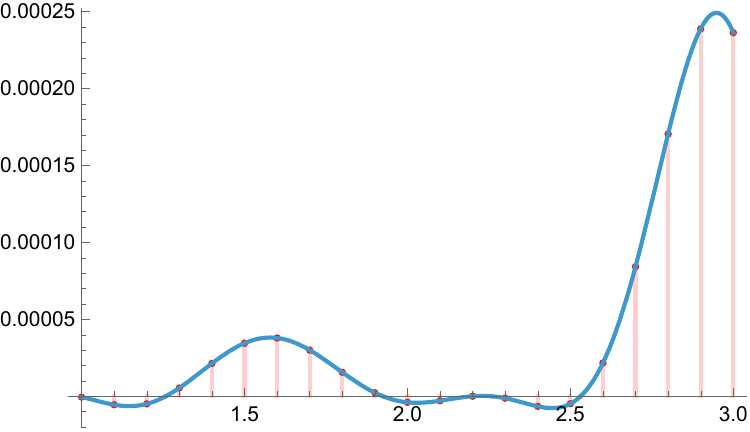}
\begin{picture}(0.3,0.4)(0,0)
    \put(5,10){\makebox(0,0){$\scriptstyle{\Ph_1}$}}
    \put(-100,100){\makebox(0,0){$\scriptstyle{\Re \mathfrak{N}_+(\Ph_1)}$}}
\end{picture}
\end{subfigure} 
\vspace{1cm}
\begin{subfigure}[a]{0.5\linewidth}
\centering
    \includegraphics[width=0.9\textwidth]{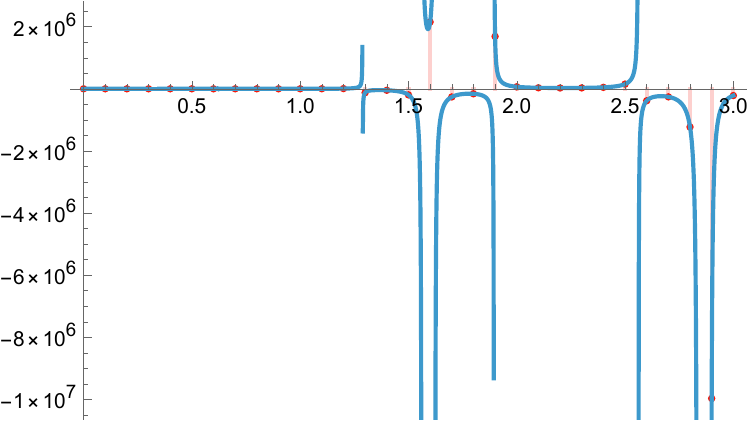}
\begin{picture}(0.3,0.4)(0,0)
    \put(5,10){\makebox(0,0){$\scriptstyle{\Ph_1}$}}
    \put(-100,120){\makebox(0,0){$\scriptstyle{\Re \mathfrak{N}_-(\Ph_1)}$}}
\end{picture}
\end{subfigure}
\begin{subfigure}[a]{0.5\linewidth}
\centering
    \includegraphics[width=0.9\textwidth]{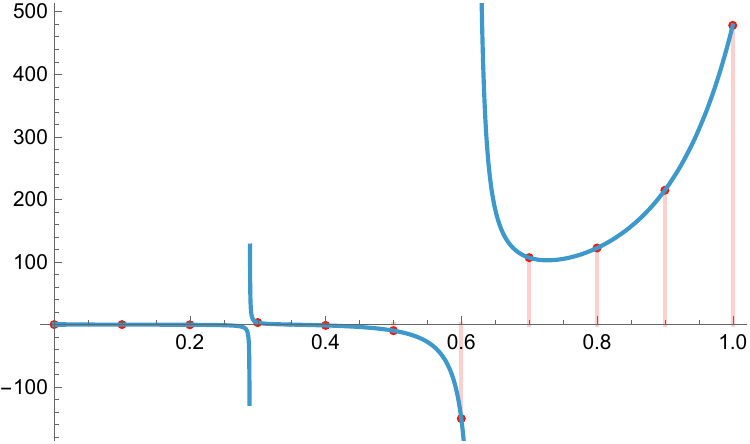}
\begin{picture}(0.3,0.4)(0,0)
    \put(5,10){\makebox(0,0){$\scriptstyle{\Ph_1}$}}
    \put(-100,100){\makebox(0,0){$\scriptstyle{\Re \mathfrak{N}_-(\Ph_1)}$}}
\end{picture}
\end{subfigure}
\caption{Numerical verification of the functional relations for the ratios of the timelike structure constants for $\bh = \frac{\pi}{2}$, i.e., for $\ch= -1.11808$. The top and bottom rows show the real part of the ratio of NS sector correlators $\mathfrak{N}_\pm(\Ph_1, \frac{1}{2}, \frac{1}{6})$, respectively. Since the function has a large range of variation, we have plotted both an extended range, and a zoomed-in version alongside. There is minimal variation in the imaginary part of the ratio, which is therefore not shown. The other conventions are as indicated in the caption of~\cref{fig:slrecursioncheck}.  }
\label{fig:tlNS-recursioncheck}
\end{figure}

\begin{figure}[htp!]
\begin{subfigure}[a]{0.5\linewidth}
\centering
    \includegraphics[width=0.9\textwidth]{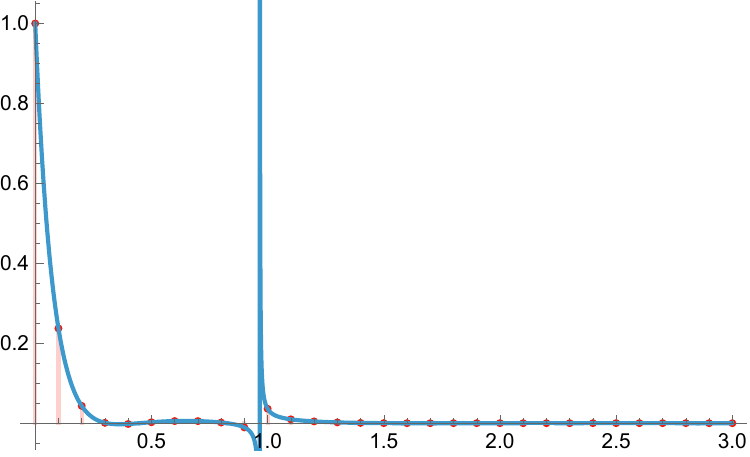}
\begin{picture}(0.3,0.4)(0,0)
    \put(5,10){\makebox(0,0){$\scriptstyle{\Ph_1}$}}
    \put(-100,100){\makebox(0,0){$\scriptstyle{\Re \mathfrak{R}_+(\Ph_1)}$}}
\end{picture}
\end{subfigure}
\begin{subfigure}[a]{0.5\linewidth}
\centering
    \includegraphics[width=0.9\textwidth]{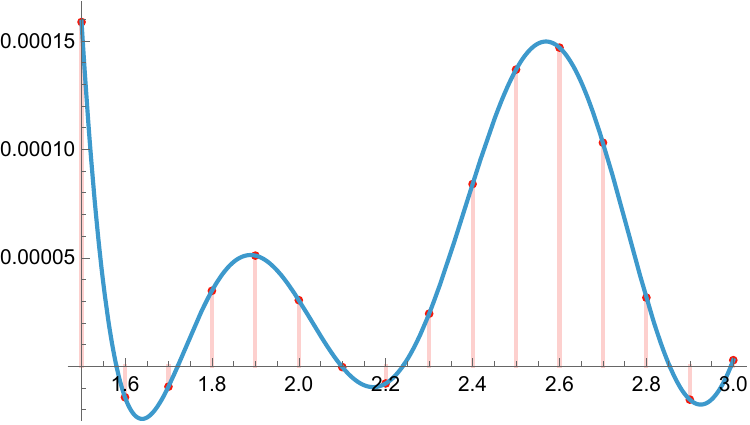}
\begin{picture}(0.3,0.4)(0,0)
    \put(5,10){\makebox(0,0){$\scriptstyle{\Ph_1}$}}
    \put(-100,100){\makebox(0,0){$\scriptstyle{\Re \mathfrak{R}_+(\Ph_1)}$}}
\end{picture}
\end{subfigure}
\vspace{1cm}
\begin{subfigure}[a]{0.5\linewidth}
\centering
    \includegraphics[width=0.9\textwidth]{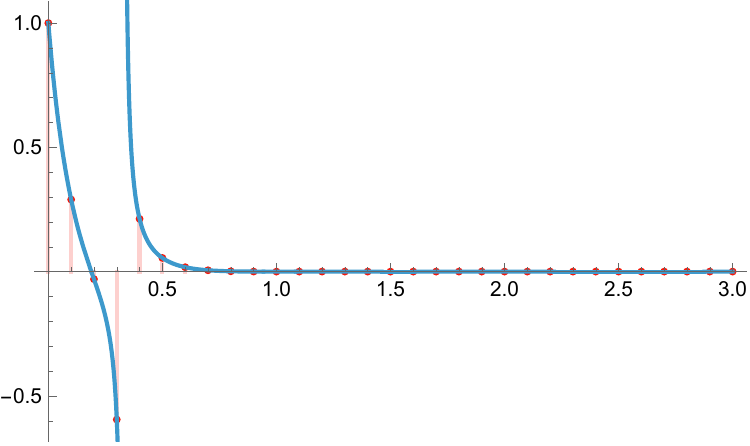}
\begin{picture}(0.3,0.4)(0,0)
    \put(5,40){\makebox(0,0){$\scriptstyle{\Ph_1}$}}
    \put(-100,100){\makebox(0,0){$\scriptstyle{\Re \mathfrak{R}_-(\Ph_1)}$}}
\end{picture}
\end{subfigure}
\begin{subfigure}[a]{0.5\linewidth}
\centering
    \includegraphics[width=0.9\textwidth]{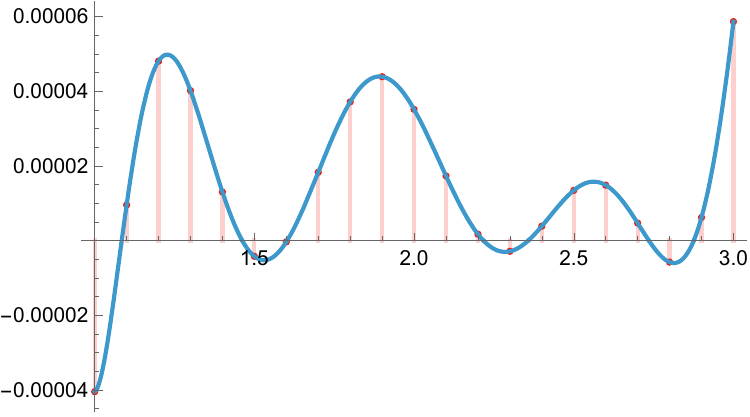}
\begin{picture}(0.3,0.4)(0,0)
    \put(5,40){\makebox(0,0){$\scriptstyle{\Ph_1}$}}
    \put(-100,100){\makebox(0,0){$\scriptstyle{\Re \mathfrak{R}_-(\Ph_1)}$}}
\end{picture}
\end{subfigure}
\caption{A numerical verification of the functional relations for the ratios of structure constants for $\bh = \frac{\pi}{2}$, i.e., for $\ch= -1.11808$. The top and bottom rows show the real part of the ratio of Ramond sector correlators $\mathfrak{R}_\pm(\frac{1}{\sqrt{2}},\Ph_2, \frac{1}{\sqrt{3}})$, respectively. Once again, these functions have a large range of variation, and so we have plotted both an extended range, and a zoomed-in version alongside. Likewise, there is minimal variation in the imaginary part of the ratio, which is therefore not shown. The conventions are otherwise as specified in~\cref{fig:slrecursioncheck}.  }
\label{fig:tlR-recursioncheck}
\end{figure}

To convince ourselves that these are the right structure constants, we first verify that they satisfy the recursion relations derived earlier~\eqref{eq:CWCVratios} and~\eqref{eq:CeCoratios}. One can, for instance, do so numerically by estimating the ratios of the  normalization independent part of the structure constants. 
We illustrate this for a representative case in~\cref{fig:tlNS-recursioncheck} and~\cref{fig:tlR-recursioncheck} for the NS and R sector structure constants, respectively.


\paragraph{An analytic verification of recursion relations:}
One can, of course, derive our answer, by picking a suitable ansatz for the solution of the functional relations. The logic is essentially to pick an analytic continuation of the special functions built from $\dGb{z}$ and verify that it does the job. For completeness, we outline the essential steps below. Readers who are convinced about the result are invited to skip ahead to our check of crossing symmetry at the end of this section (or directly to~\cref{sec:svms}). 

Let us first consider the analytic continuation of the normalization independent part of the structure constants. Focus on the NS sector data characterized by $\CVh{\bh}$ and $\CWh{\bh}$. From the analysis of degenerate correlators we have our functional relation~\eqref{eq:CWCVratios}, which should remain valid in the timelike case.  Rather than derive the results directly as quoted, we will take a detour by introducing specific combinations of the double-Gamma function. This allows us to parallel the DOZZ type derivation of the structure constants. 

Recall that the Upsilon function appearing in the traditional presentation of the DOZZ
formula is given by the combination
\begin{equation}\label{eq:upsb}
\Ub{z} = \frac{1}{\dGb{z}\, \dGb{Q-z}}\,.
\end{equation}
We now introduce two combinations, $\UNS{b}{z}$ and $\UR{b}{z}$, analogously to $\dGNS{b}{z}$ and $\dGR{b}{z}$, defining them via products of $\Ub{z}$ with some shifted arguments. Specifically,  
\begin{equation}\label{eq:upsNSR}
\begin{split}
\UNS{b}{z} = \Ub{\frac{z}{2}}\, \Ub{\frac{z+Q}{2}}\,, \qquad
\UR{b}{z} = \Ub{\frac{z+b}{2}}\, \Ub{\frac{z+b^{-1}}{2}}\,.
\end{split}
\end{equation}
These functions inherit functional relations from the double-Gamma function, cf.~\eqref{eq:Ubfrel} and~\eqref{eq:UNSRfrel}. The expression for the spacelike structure constants in terms of these functions can be found in~\cref{sec:norms}. As described there, the main difference is that we chose to normalize the vertex operators differently. 

The analytic continuation of the Upsilon function to imaginary values of the Liouville parameter, defines a new set of functions, which we indicate with a hat decoration apart from changing to sans serif font for ease of visualization (they will only appear in this subsection). The analytic continuation is defined as 
\begin{equation}\label{eq:Uhdefs}
\UNSh{b_*}{z} =\frac{1}{\UNS{ib_*}{-i\,z+i\,b_*}}\,, \qquad 
\URh{b_*}{z} =\frac{1}{\UR{ib_*}{-i\,z+i\,b_*}} \,.
\end{equation}
Here we introduced $b_* \in i\, \mathbb{R}_-$, to make clear that the r.h.s.\ is defined in terms of the standard Upsilon functions. 
The l.h.s.\ defines functions that parameterize the analytically continued theory.  We could have traded it for the timelike Liouville parameter $\bh$, which is obtained by rotating $i\,b_*\to \bh \in \mathbb{R}_+$. We refrain from doing so, and will write all the expressions in this subsection in terms of $b_*$. 
Notice that in the analytic continuation, the argument of the Upsilon function is rotated the opposite way,  as is evident from  the r.h.s.\ of~\eqref{eq:Uhdefs}.
 
With this continuation, the new set of functions satisfy functional relations inherited from those of the double-Gamma function~\eqref{eq:dGfrel}. For instance, by adapting~\eqref{eq:UNSRfrel} one can  verify that  
\begin{equation}\label{eq:Uhfrel}
\begin{split}
\frac{\UNSh{b_*}{z+b_*}}{\URh{b_*}{z}} 
&= (i\,b_*)^{1-b_*\,z} \, \gamma\pqty{\frac{b_*\,z}{2}}\,, \\ 
\frac{\URh{b_*}{z+b_*}}{\UNSh{b_*}{z}} 
&= (i\,b_*)^{-b_*\,z} \, \gamma\pqty{\frac{1+b_*\,z}{2}}\, \,.
\end{split}
\end{equation}

We now guess a solution to the relations~\eqref{eq:CWCVratios} for the timelike structure constants. We take as our anstaz 
\begin{equation}\label{eq:tLCWCVguess}
\begin{split}
\CVh{\bh}(\Ph_1,\Ph_2,\Ph_3)
&=
    \pqty{\Prod{i=1}{3}\, \mathsf{N}_{_\mathrm{NS}}(\Ph_i)}
    \Prod{\epsilon_{2,3} = \pm 1}{}\,
    \frac{1}{\URh{b_*}{\frac{b_* + b_*^{-1}}{2} +i\,\Ph_1 +i\, \epsilon_2 \, \Ph_2 + i\,\epsilon_3\,\Ph_3}}\\ 
\CWh{\bh}(\Ph_1,\Ph_2,\Ph_3)
&=
   2i\, \etaW\, \pqty{\Prod{i=1}{3}\, \mathsf{N}_{_\mathrm{NS}}(\Ph_i)}
    \Prod{\epsilon_{2,3} = \pm 1}{}\,
    \frac{1}{\UNSh{b_*}{\frac{b_* + b_*^{-1}}{2} + i\,\Ph_1 + i\,\epsilon_2 \, \Ph_2 + i\, \epsilon_3\,\Ph_3}}\,, 
\end{split}
\end{equation}
All the quantities in the r.h.s.\ are defined in terms of $b_*$, which we recall can be traded for $\bh$, and the momenta $\Ph_i$. 

It remains to verify that these functions solve the recursion relation. We can do this by computing the ratio of the structure constants with $\Ph_1$ shifted by $i\epsilon\,\frac{b_*}{2}$ using~\eqref{eq:Uhfrel}. For example, consider the ratio of the normalization independent part of the NS structure constants as before, viz.,  
\begin{equation}
\begin{split}
\mathfrak{N}_\epsilon(\Ph_1,\Ph_2,\Ph_3) 
&\equiv 
\frac{\CWh{\bh}(\Ph_1 + \epsilon\, \frac{i\,b_*}{2},\Ph_2, \Ph_3)}{\CVh{\bh}(\Ph_1 -\epsilon\, \frac{i\,b_*}{2},\Ph_2, \Ph_3)} \Bigg|_{\text{\st{norm}}} \\
&= 
 \Prod{\epsilon_{2,3} = \pm 1}{}
\frac{\URh{b}{\frac{b_* + b_*^{-1}}{2} + i\,\Ph_1  + \frac{\epsilon\,b_*}{2}+ i\, \epsilon_2\, \Ph_2 + i\,\epsilon_3\, \Ph_3}}{\UNSh{b}{\frac{b_* + b_*^{-1}}{2} + i\,\Ph_1 -\frac{\epsilon\,b}{2} +i\, \epsilon_2\, \Ph_2 + i\,\epsilon_3\, \Ph_3}} \,.
\end{split}
\end{equation}
The r.h.s.\ can be expressed in terms of familiar Upsilon functions using our definition for the analytic continuation~\eqref{eq:Uhdefs}. We can, however, directly compute the ratio using~\eqref{eq:Uhfrel}   to obtain
\begin{equation}\label{eq:tLCWCVcheck1}
\begin{split} 
\mathfrak{N}_\epsilon(\Ph_1,\Ph_2,\Ph_3) 
&= 
\Prod{\epsilon_{2,3} = \pm 1}{}\, 
(i\,b_*)^{-\delta_{\epsilon,-1}-i\,\epsilon\,b_*\pqty{\frac{Q_*-b_*}{2} + i\,\Ph_1 + i\,\epsilon_2\, \Ph_2 + i\,\epsilon_3\, \Ph_3}} \\
&\qquad \times
\gamma\pqty{\frac{\delta_{\epsilon,1}}{2} + \frac{b}{2}\, 
\pqty{\frac{Q_*-b_*}{2}+  i\,\Ph_1 + i\,\epsilon_2\, \Ph_2 +i\, \epsilon_3\, \Ph_3}}\\
&\qquad = 
    (i\,b_*)^{-2-4\,i\,\epsilon\,b_*\, \Ph_{1}}
    \Prod{\epsilon_{2,3} = \pm 1}{}\,
    \gamma\pqty{\frac{3}{4}+i\,\frac{b_*}{2}\, \pqty{\epsilon\,\Ph_1 + \epsilon_2\, \Ph_2 + \epsilon_3\, \Ph_3}} \,,
\end{split}
\end{equation}
which is the desired answer. We introduced $Q_* \equiv b_*  + b_*^{-1}$ in the intermediate step.

The final thing we need to do is furnish normalization factors satisfying the recursion relation. 
In the spacelike super Liouville theory, the normalization of the structure constants in the conventional presentation is proportional to $N_s(P) = \UNS{b}{Q+2\,i\,P}$, see~\eqref{eq:NPdef} for the precise definition. This part of the normalization factor therefore satisfies the functional relation 
\begin{equation}
\frac{N_s\pqty{P+\frac{i\,b}{2}}}{N_s\pqty{P-\frac{i\,b}{2}}} =  b^{1+b^2+ 4\,i\, b\, P}\,\gamma\pqty{\frac{1}{2}-i\,b\,P} \, \gamma\pqty{-\frac{b^2}{2}-i\,b\,P} \,.
\end{equation}
Likewise, it is easy to verify that the choice 
\begin{equation}
\mathsf{N}_t(\Ph) = \URh{b_*}{Q_*+ 2\,i\,\Ph} \,,
\end{equation}
continues to satisfy a similar relation. For instance, 
\begin{equation}
\begin{split}
\frac{\mathsf{N}_t(\Ph + i\,\frac{b_*}{2})}{\mathsf{N}_t(\Ph-i\,\frac{b_*}{2})}
&=
    \frac{\URh{b_*}{Q_*+2\,i\,\Ph-b_*}}{\URh{b_*}{Q_* +2\,i\,\Ph+b_*}} 
= 
    \frac{\UR{ib_*}{i\,b_*-i\,(Q_*+2\,i\, \Ph+ b_*)}}{\UR{ib_*}{i\,b_*-i\,(Q_*+2\,i\,\Ph-b_*)}}\\ 
&= 
    \frac{\Upsilon_{ib_*}\left(\frac{i\,b_*-i\,(Q_*+2\,i\Ph + b_*)}{2}\right)\,
    \Upsilon_{ib_*}\left(\frac{-i\,b_*^{-1}-i\,(Q_*+2\,i\,\Ph)}{2}\right)}{\Upsilon_{ib_*}\left(\frac{i\,b_*-i\,(Q_*+2\,i\,\Ph)}{2}+i\,b_*\right)\,
    \Upsilon_{i\,b_*}\left(\frac{-i\,b_*^{-1}-i\,(Q_*+2\,i\,\Ph)}{2}+i\,b_*\right)}  \\ 
&=
    (ib_*)^{1+b_*^2+4i\,b_*\,\Ph}\,\gamma\left(\frac{1}{2}-i\,b_*\,\Ph\right)\gamma\left(-i\,b_*\,\Ph-\frac{b_*^2}{2}\right).
\end{split}
\end{equation} 
We therefore have a solution to the recursion relation that allows us to define the timelike super Liouville theory in the desired range $c< \frac{3}{2}$. One can improve $\mathsf{N}_t(\Ph)$ and present it in a form similar to the one  used in the spacelike case, i.e., the quantity referred to as $\NNS(P)$ in~\eqref{eq:NPdef}.  The analog for the timelike theory is 
\begin{equation}\label{eq:NPhdef}
\mathsf{N}_{_\mathrm{NS}}(\Ph) =
\pqty{\URo\,\bqty{ \pi\,\mu\, (ib_*)^{1-b_*^2}\,
    \gamma\pqty{\frac{b_*\,Q_*}{2}} }
  ^{-\frac{Q_*+3\,i\,\Ph}{2\,b_*} }}^{\frac{1}{3}}\, \URh{b_*}{Q_* + 2\,i\,\Ph}\,.
\end{equation}
The analysis of correlators with Ramond operators proceeds similarly. The reader can repurpose the formulae from~\cref{sec:norms} to arrive at analogous expressions for $\Ceh{\bh}{}$ and $\Coh{\bh}{}$. 

Recall that in the spacelike case, we have chosen to work with a normalization inspired by the Cardy density of states. While we cannot use the same logic here, we can, of course, modify the normalization of the operators. 
In presenting our results in~\eqref{eq:tLCs} we have made a particular choice inspired by the application to the worldsheet string construction we describe in~\cref{sec:svms}. As we shall see later, with this choice there are simple relations between products of the spacelike and timelike structure constants, with suitable identifications of parameters. In addition, while writing those expressions, we have avoided introducing new special functions, and presented the answer in terms of the double-Gamma function directly to keep the notation light. In doing so, we accounted for the analytic continuation of the momenta and replaced $i\,b_*$ by $\bh$ (specifically, $\Ph,\bh\in\mathbb{R}_+$).

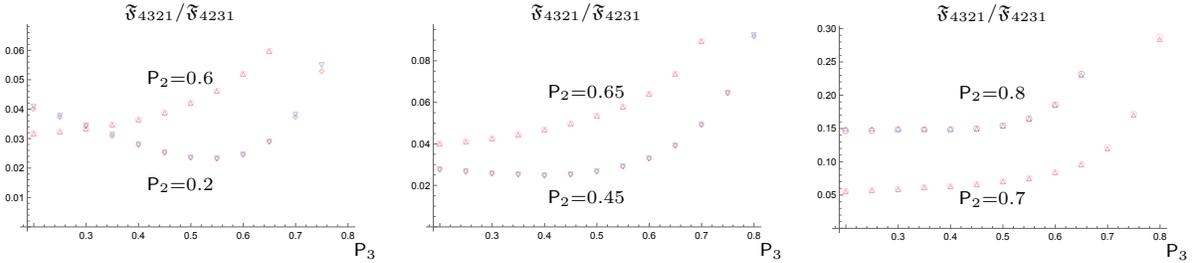
\begin{figure}[htp!]
\begin{subfigure}[a]{0.3\linewidth}
\centering
\includegraphics[width=\textwidth]{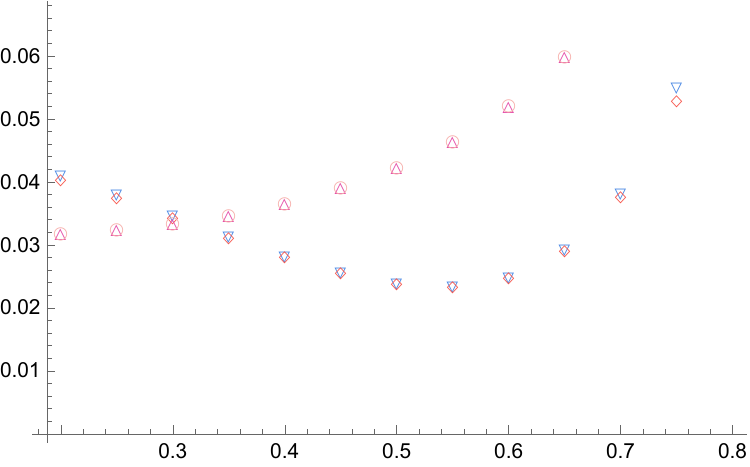}
\begin{picture}(0.3,0.4)(0,0)
\put(70,10){\makebox(0,0){$\scriptstyle{\Ph_3}$}}
\put(0,100){\makebox(0,0){$\scriptstyle{\mathfrak{F}_{4321}/\mathfrak{F}_{4231}}$}}
\put(0,35){\makebox(0,0){$\scriptstyle{\Ph_2 =0.2}$}}
\put(0,75){\makebox(0,0){$\scriptstyle{\Ph_2 =0.6}$}}
\end{picture}
\end{subfigure}
\hspace{5mm}
\begin{subfigure}[a]{0.3\linewidth}
\centering
\includegraphics[width=\textwidth]{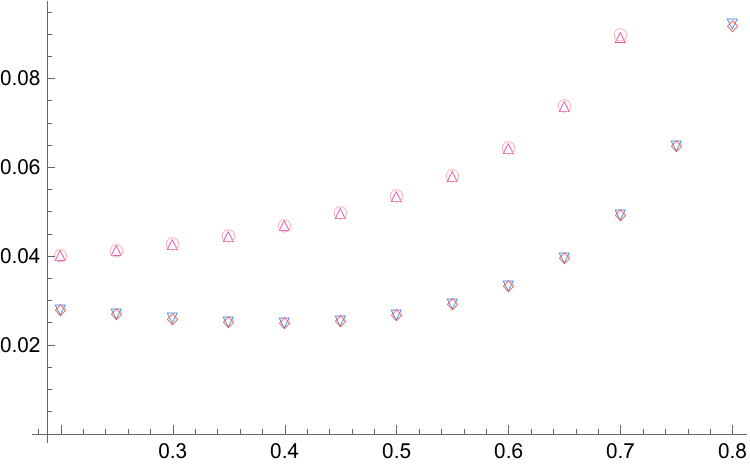}
\begin{picture}(0.3,0.4)(0,0)
\put(70,10){\makebox(0,0){$\scriptstyle{\Ph_3}$}}
\put(0,100){\makebox(0,0){$\scriptstyle{\mathfrak{F}_{4321}/\mathfrak{F}_{4231}}$}}
\put(0,30){\makebox(0,0){$\scriptstyle{\Ph_2 =0.45}$}}
\put(0,70){\makebox(0,0){$\scriptstyle{\Ph_2 =0.65}$}}
\end{picture}
\end{subfigure}
\hspace{5mm}
\begin{subfigure}[a]{0.3\linewidth}
\centering
\includegraphics[width=\textwidth]{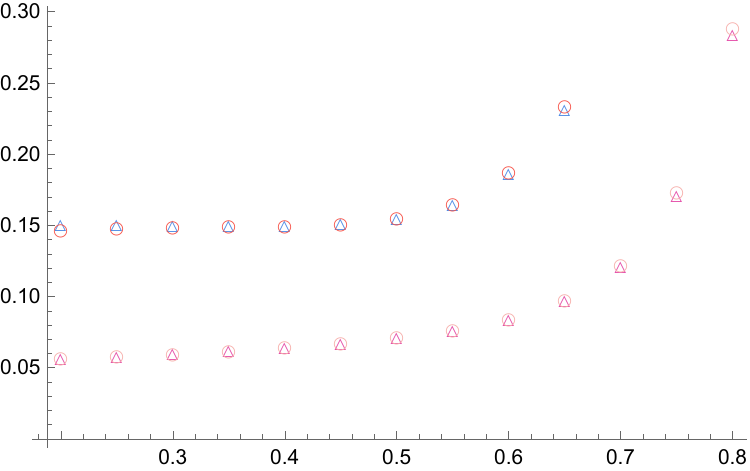}
\begin{picture}(0.3,0.4)(0,0)
\put(70,10){\makebox(0,0){$\scriptstyle{\Ph_3}$}}
\put(0,100){\makebox(0,0){$\scriptstyle{\mathfrak{F}_{4321}/\mathfrak{F}_{4231}}$}}
\put(0,30){\makebox(0,0){$\scriptstyle{\Ph_2 =0.7}$}}
\put(0,70){\makebox(0,0){$\scriptstyle{\Ph_2 =0.8}$}}
\end{picture}
\end{subfigure}
\caption{ Check of crossing symmetry of the NS sector 4-point function $\mathfrak{F}_{4321}(z)$ defined in~\eqref{eq:F4fn}. We set $\bh =1$ and  thus $\ch = \frac{3}{2}$.  We fixed two external momenta $\Ph_1 = 0.5$, $\Ph_4 = 0.6$, along with the cross-ratio, $z= \frac{1}{3} + \frac{i}{2}$, and scanned over different values of $\Ph_3,\Ph_4 \in [0.2,0.8]$ in steps of $0.05$.  The plots show the results for the direct and cross channel correlator as a function of $\Ph_3 \in [0.2,0.8]$. Six different values of $\Ph_2$ shown as indicated. We emphasize that there are two overlapping markers for each value of $\Ph_{2,3}$ in the plots, which indicates that the results are crossing invariant, with small errors that are barely discernable (for the most part) to the naked eye. Numerical values are tabulated in~\cref{tab:numcrosscheck}  for comparison.}
\label{fig:NSb1cross}
\end{figure}

 \paragraph{Numerical verification of crossing symmetry:} Another consistency check one can undertake is to show that the structure constants derived for the timelike theory satisfy crossing symmetry. This can be done by evaluating the 4-point function of NS and R operators in the direct and cross channel and comparing the result. We illustrate excellent matching for a representative example in~\cref{fig:NSb1cross}. We also compiled some sample values of the correlators in~\cref{tab:numcrosscheck} to illustrate the numerical agreement.

To explain the result, consider the 4-point function
\begin{equation}\label{eq:F4fn}
\mathfrak{F}_{4321}(z) = \expval{\Vh_{\Ph_4}(\infty)\, \Wh_{\Ph_3}(1)\, \Wh_{\Ph_2}(z)\, \Vh_{\Ph_1}(0)} \,.
\end{equation}
We examine the crossing relation arising from exchanging the operators labeled by $\Ph_2$ and $\Ph_3$, which demands
that the direct channel  (given by the ordering $\Ph_4\Ph_3\Ph_2\Ph_1$) answer be related to the cross channel (the ordering $\Ph_4\Ph_2\Ph_3\Ph_1$) result, viz.,
\begin{equation}
\mathfrak{F}_{4321}(z) = \abs{z}^{2\,(\hh_4 - \hh_3 - \hh_2 - \hh_1 -1)} \, \mathfrak{F}_{4231}(z^{-1}) \,.   
\end{equation}

This direct channel correlator can be expressed in terms of the structure constants and the 4-point superconformal blocks, as 
\begin{equation}
\begin{split}
\mathfrak{F}_{4321}(z)
&=
\int_0^\infty\, \frac{d\Ph}{\pi}\, \frac{(i\,\Ph)^2}{\rhoNS(i\,\Ph)}
\Bigg[
\CWh{\bh}(\Ph_3,\Ph_4, \Ph+i\,\varepsilon_\Ph)\, \CWh{\bh}(\Ph_2,\Ph_1,-\Ph-i\,\varepsilon_\Ph) \,\abs{\mathcal{F}^{\mathrm{e}}(\hh_4,\hh_3, \hh_2,\hh_1; \hh|z)}^2 \\ 
&\qquad \qquad \quad 
-\CVh{\bh}(\Ph_4, \Ph_3, \Ph+i\,\varepsilon_\Ph)\, \CVh{\bh}(\Ph_2,\Ph_1,-\Ph-i\,\varepsilon_\Ph) \,\abs{\mathcal{F}^{\mathrm{o}}(\hh_4,\hh_3, \hh_2,\hh_1; \hh|z)}^2 
\Bigg].
\end{split}
\end{equation}
The superconformal blocks $\mathcal{F}^\mathrm{e/o}$ depend on the external weights $\hh_i$ determined from $\Ph_i$ and also on the internal weight $\hh = \frac{1}{2} \pqty{-\frac{1}{4}\,\Qh^2 + \Ph^2}$. The superscripts indicate `even/odd' blocks which arise depending on whether the internal operator is an integer or half integer level descendant. 
We summarize the essential features of the superconformal blocks in~\cref{sec:superrecurion}. The point of import is that one can efficiently compute them using recursion. A similar expression can be derived for the cross channel correlator. Evaluating both integrands using the recursion formulae and our prediction for the structure constants, and furthermore performing the integral over the internal weight, we can ascertain whether they agree. In performing the numerical integral, we have to shift the contour of the internal Liouville momentum away from the real axis, which is indicated by the $i\,\varepsilon_{\Ph}$  in the above expression, as explained earlier.
For the range of values analyzed, it proved sufficient to take $\varepsilon_{\Ph} =0.08$. 
\begin{table}
    \centering
    \begin{tabular}{|c|c||c | c|}
    \hline 
     $\{\Ph_4,\Ph_3,\Ph_2,\Ph_1\}$    &  $z$  & $\mathfrak{F}_{4321}(z)$  & $\mathfrak{F}_{4231}(z^{-1})$   \\ 
     \hline 
     $\{0.6, 0.4, 0.333, 0.5\}$   & 
     $\frac{1}{3} + \frac{i}{2}$ & 
     $0.0253334 -1.58815 \times 10^{-9}\, i $ &  
     $0.0253012 -1.31516 \times 10^{-8}\,i $  \\
     $\{0.6,0.5,0.333,0.5\}$   & 
     $\frac{1}{3} + \frac{i}{2}$ & 
     $0.0237654 -4.62706 \times 10^{-10}\, i $ &  
     $0.0237708 -1.55996 \times 10^{-9}\,i $  \\
      $\{0.6,0.6, 0.333,0.5\}$   & 
     $\frac{1}{3} + \frac{i}{2}$ & 
     $ 0.0274788 -1.13394 \times 10^{-10}\, i $ &  
     $ 0.0274931 -9.36453 \times 10^{-10}\,i $  \\
      $\{0.6,0.4,0.333,0.5\}$   & 
     $\frac{1}{2} + \frac{i}{7}$ & 
     $0.0563771 + 2.13843 \times 10^{-9}\, i $ &  
     $0.0559575 -1.23735 \times 10^{-7}\,i $  \\
      $\{0.6,0.5, 0.333,0.5\}$   & 
     $\frac{1}{2} + \frac{i}{7}$ & 
     $0.0572776 +7.94466 \times 10^{-10}\, i $ &  
     $0.0572766 -1.94196 \times 10^{-8}\,i $  \\
    $\{0.6,0.6, 0.333,0.5\}$   & 
     $\frac{1}{2} + \frac{i}{7}$ & 
     $0.0704186 -3.3318 \times 10^{-10}\, i $ &  
     $0.0699274 -4.86935 \times 10^{-8}\,i $  \\
     \hline
    \end{tabular}
    \caption{Numerical values of the direct and cross channel correlator to illustrate crossing symmetry for $\bh =1$. The offset for the internal momentum was chosen as $\varepsilon_{\Ph} = 0.08$. We have indicated the imaginary part of the answer for illustrative purpose. }
    \label{tab:numcrosscheck}
\end{table}

While we only checked crossing for one value of the central charge $\ch =1$, and that too only for one of the NS sector correlators, the numerical agreement strongly supports our prediction for the structure constants.\footnote{ Victor Rodriguez has independently checked that crossing continues to hold for other values of $\bh$. We thank him for sharing his results with us.} It should be possible to similarly check other correlators, including those with Ramond operators, using the elliptic recursion formulae for the superconformal blocks given  in~\cref{sec:superrecurion}.   

\section{The super Virasoro minimal string}\label{sec:svms}

Having defined the timelike super Liouville theory, we can now proceed to construct a worldsheet string theory using it and the spacelike super Liouville theory as matter SCFTs. We couple these to worldsheet $\mathcal{N} =1$ supergravity, and gauge the superdiffeomorphisms and super-Weyl symmetries. 

One can characterize the worldsheet theory in terms of the Liouville SCFTs and the ghost CFTs. To wit, the ingredients are  
\begin{enumerate}
\item A spacelike super Liouville theory with $c=\frac{3}{2}+3\,Q^2\geq \frac{27}{2}$. We set $Q=b+b^{-1}$, with $b\in \mathbb{R}_+$.
\item  A timelike super Liouville theory with $\ch=\frac{3}{2}-3\,\Qh^2 \leq\frac{3}{2}$. We let $\Qh=\bh^{-1} - \bh$, with $\bh\in \mathbb{R}_+$.
\item A $\mathfrak{b}\mathfrak{c}-\beta\gamma$ ghost SCFT corresponding to the worldsheet superdiffeomorphisms, with central charge $c_\mathrm{gh} = -26 + 11 = -15$.
\end{enumerate}

The worldsheet theory is anomaly free provided $c + \ch + c_\mathrm{gh} = 0$. This demands that the two Liouville theories are parameterized by a single parameter, which we  take to be $b$. Specifically, we require 
\begin{equation}\label{eq:QQhrel}
Q^2 - \Qh^2 = 4\,, \qquad \bh = b\,.
\end{equation}

We will examine the basic features of this worldsheet theory, arguing first that it provides a one-parameter generalization of the JT supergravity by examining the semiclassical limit. We then analyze the worldsheet spectrum and 3-point amplitudes and make brief contact with the matrix model results described in~\cref{sec:intro}.

\subsection{The classical limit: JT supergravity}\label{sec:JTsugra}

To relate the worldsheet string to JT supergravity, let us start with the action for the
latter, which has a compact form in superspace~\cite{Chamseddine:1991fg} (see also~\cite{Forste:2017kwy}). The observation we sketch
below was presented earlier in~\cite{Fan:2021wsb}, where the reader can find further details.

Using our conventions for superspace coordinates from~\cref{sec:spL}, the supergravity fields are a packaged into a super-tetrad $\SF{E}\indices{^A_B}$ and its super-connection. Here $\{A,B\}$ etc., are the
local frame indices with the convention $A = (a, \alpha)$ where $a$ is the usual Grassmann even frame index, and $\alpha$ the Grassmann odd frame index. The components are simple to state in the Wess-Zumino gauge: they comprise the bosonic tetrad $e^a_\mu$, the gravitino $\eta^{\alpha}_\mu$, an auxiliary field $A$, where $\mu$ indexes the coordinate basis. In addition, one has a dilaton superfield which we denote as $\mathring{\Phi}_{\mathrm{JT}}$.

For presenting the action, it will suffice to write down the component form of the superdeterminant $\SF{E} \equiv \mathrm{sdet}(\SF{E}\indices{^A_B}) $
and the supercurvature $\SF{R}_{+-}$, which are 
\begin{equation}
\begin{split}
\SF{E}
 & =
e\, \bqty{ 1 - \frac{1}{4}\, \thetab\,\theta\, A + \text{fermions}  }\,, \\
\SF{R}_{+-}
 & = A + \frac{1}{2}\, \thetab\,\theta\, (R + \frac{1}{2}\,A^2 ) +
\text{fermions} \,.
\end{split}
\end{equation}
We dropped all the fermion terms for simplicity, and refer the reader to~\cite{Fan:2021wsb} for explicit formulae.

In superspace, the JT supergravity action has a simple form
\begin{equation}\label{eq:sJT}
\begin{split}
S_\mathrm{sJT}
 & =
\int d^2z \,d^2\theta\,  \SF{E}\, \mathring{\Phi}_{\mathrm{JT}}(\SF{R}_{+-}+2)
+2\,\int dx\,d\theta\, \mathring{\Phi}_{\mathrm{JT}}\, \SF{K} \\
 & =
\int \,d^2z\, e \, \bqty{\phi\, (R+2) + \text{gravitino terms} }
+ S_\mathrm{bdy}\,.
\end{split}
\end{equation}
In the second line we have indicated the result of superspace integration in a familiar form. However, to make contact with the super Liouville theory, it will be simplest to work with the superspace integrand. 
All we will need is to exploit the fact that in superconformal gauge  one can characterize the geometric curvature data in terms of a superfield $\SF{G}$, viz.,
\begin{equation}
\SF{R}_{+-} = 2\, e^{-\SF{G}}\, D\overline{D} \SF{G}\,,
\qquad
\SF{E} = e^{\SF{G}}\,.
\end{equation}

On the other hand, in superconformal gauge, the super Virasoro minimal string
worldsheet action just has the two super Liouville theories, i.e.,
\begin{equation}
S_\mathrm{ws} =
\frac{1}{2\pi}
\int d^2 z\, d^2 \theta
\left( D \mathring{\Phi}\, \overline{D} \mathring{\Phi}
+ 4\pi i\, \mu\,  e^{b\,\mathring{\Phi}} \right) -
\frac{1}{2\pi}\int d^2z\,d^2\theta
\left(D\mathring{\Xi}\, \overline{D}\mathring{\Xi}
+ 4\pi\, i\, \mu\,  e^{b\, \mathring{\Xi}}\right) .
\end{equation}
Now consider the field redefinition
\begin{equation}
\mathring{\Phi} =
\frac{1}{b}\, \SF{G} -\pi \, b\, \mathring{\Phi}_\mathrm{JT}\,,
\qquad
\mathring{\Xi} = \frac{1}{b}\, \SF{G}
+ \pi \, b\, \mathring{\Phi}_\mathrm{JT}\,.
\end{equation}
The worldsheet action can be written in these variables as
\begin{equation}
S_\mathrm{ws}
= -2\int d^2z\,d^2\theta\, \left[
  D\SF{G} \,\overline{D}\mathring{\Phi}_\mathrm{JT}
  +2\, i\,\mu\,  e^{\SF{G}}\,
  \sinh (\pi b^2\,\mathring{\Phi}_\mathrm{JT})\right] + S_\mathrm{bdy}.
\end{equation}
Upon integrating the bulk term by parts one obtains a dilaton gravity action
\begin{equation}
S_\mathrm{ws}=
-\int d^2z\,d^2\theta \, \SF{E}\left[\mathring{\Phi}_\mathrm{JT}\,
  \SF{R}_{+-} + 4\,i\, \mu\, \sinh( \pi b^2\,\mathring{\Phi}_{\mathrm{JT}})\right] + S_\mathrm{bdy}.
\end{equation}
Finally, setting
\begin{equation}
\mu = -\frac{i}{4\sin(\pi b^2)}\,,
\end{equation}
and taking $b\to 0$, we see that we recover the JT supergravity action.
This establishes the classical connection we sought. Our aim is, of course, to
examine the quantum theory, which we turn to next.

\subsection{The quantum string: vertex operators}\label{sec:vops}

The quantum string vertex operators are obtained as usual by imposing the Virasoro constraints. We have operators in both the NS-NS 
and R-R sectors. We expect to be able to define two theories, a Type 0A and a Type 0B theory, which are obtained by two different choices for the GSO projection. The distinction, as we will shortly see, will turn out to lie with the Ramond sector states one retains.

To define the GSO projection, we should characterize the worldsheet fermion number operator $(-1)^{F_\mathrm{ws}}$. We do so as follows. Let $(-1)^{F_s}$ and $(-1)^{F_t}$ be the fermion number operator in the spacelike and timelike super Liouville theories, respectively. In addition, let the ghost fermion number operator be $(-1)^{F_\mathrm{gh}}$. We define the matter fermion number operator ($F_M$) and the total worldsheet fermion number operator $F_\mathrm{ws}$ in terms of these as 
\begin{equation}
   (-1)^{F_M} =  (-1)^{F_s}\, (-1)^{F_t} \,, \qquad 
   (-1)^{F_\mathrm{ws}} = (-1)^{F_M} \, (-1)^{F_\mathrm{gh}} \,.
\end{equation}
The worldsheet superconformal current modes have contributions from both the Liouville theories. They are taken to be
\begin{equation}\label{eq:Gws}
    \mathcal{G}_r = G_r^\mathrm{sL} \otimes (-1)^{F_t} + \mathbb{I}^\mathrm{sL} \otimes G_r^\mathrm{tL}\,, 
    \qquad
    \widetilde{\mathcal{G}}_r = \widetilde{G}_r^\mathrm{sL} \otimes (-1)^{F_t} +  \mathbb{I}^\mathrm{sL} \otimes \widetilde{G}_r^\mathrm{tL}\,,
\end{equation}
where we use the superscripts to indicate the super Liouville origin. The above expression also includes the cocycle factors to ensure that these current satisfy the $\mathcal{N} =1$ superconformal algebra. 

\paragraph{NS vertex operators:} Let us start with the NS sector, which is common to both the 0A and 0B theories. In this case, we want to construct a worldsheet superprimary operator that satisfies the super Virasoro constraints. 

Consider, the combination of the superprimary operator of the spacelike and timelike super Liouville theories, dressed with the ghosts. The weights of $V_P$ and $\Vh_{\Ph}$ are $\hP$ and $\hhP$ given in~\eqref{eq:sLhp}  and~\eqref{eq:tLhp}. Furthermore, the ghosts give a contribution of $\frac{1}{2}$. Therefore, the Virasoro constraint implies upon using~\eqref{eq:QQhrel} a relation between the spacelike and timelike momenta, viz.,
\begin{equation}
\frac{Q^2}{8}+\frac{1}{2}\, P^2
-\frac{\Qh^2}{8}+\frac{1}{2}\, \Ph^2=\frac{1}{2} 
\;\; \Longrightarrow \; \; \Ph = -i\,P \,.
\end{equation}
This was the  choice of analytic continuation of Liouville momenta encountered earlier in defining the timelike theory. It is indeed reassuring that the same choice is consistent for the worldsheet description. 

Since we are discussing the superstring, we have to worry about the picture number arising from the $\beta\gamma$ ghost CFT. For the present, we will simply write down vertex operators in different pictures.  In the $(-1,-1)$ picture, the unintegrated physical vertex operator is therefore
\begin{equation}\label{eq:NSvo11}
\mathcal{V}_P^{(-1,-1)}
=
    g_s\, \mathfrak{c}\, \tilde{\mathfrak{c}}
    \, e^{-\varphi-\tilde{\varphi}}
    \, V_P\, \Vh_{\Ph=-i\,P}\,.
\end{equation}
Here $e^{-\varphi}$ and $e^{-\tilde{\varphi}}$ are the holomorphic and anti-holomorphic bosons of the $\beta\gamma$ CFT. 
We will later need the representative in the $(0,0)$ picture as well. 
\begin{equation}\label{eq:NSvo00}
\begin{split}
\mathcal{V}_P^{(0,0)}
&=
    g_s\, \mathfrak{c}\, \tilde{\mathfrak{c}}\, 
    \,\mathcal{G}_{-\frac{1}{2}} \tilde{\mathcal{G}}_{-\frac{1}{2}} (V_P\, \Vh_{\Ph=-i\,P})\,, \\
&=
    g_s\, \mathfrak{c}\, \tilde{\mathfrak{c}}\, 
    \pqty{\Lambda_P\, \tilde{\Lambdah}_{-i\,P}
    + \tilde{\Lambda}_P\, \Lambdah_{-i\,P} 
    - V_P\, \Wh_{-i\,P} - W_P\, \Vh_{-i\,P}} \,.    
\end{split}    
\end{equation}
The action of the supercurrent mode was deduced from the superfield expansions~\eqref{eq:Snsop}
and~\eqref{eq:Tnsop}. Generally, we will have an admixture of contributions from the  operators with non-zero spin, viz., terms like 
$\Lambda_P\, \tilde{\Lambdah}_{\Ph}$ part from pieces that involve spinless operators from both theories, e.g., $V_P\, \Wh_{\Ph}$ and $W_P\, \Vh_{\Ph}$, as indicated in the equation.

\paragraph{R-R vertex operators:}
To understand the R-R operators, we will first examine the states in the Ramond sector for the spacelike and timelike theories separately. Subsequently, we can put them together and implement the GSO projection. 

In the Ramond sector, we work with vertex operators $\RsL{\pm}{P}$ which have definite fermion number. The super Liouville theories have a single fermion number operator $(-1)^{F_s}$ and $(-1)^{F_t}$, respectively. We have picked the two vertex operators to be eigenstates of these operators in the two theories. 

In the spacelike case, the OPE of the supercurrent with the Ramond vertex operators $\RsL{\pm}{P}$, which carry definite fermion number, is given by   
\begin{equation}
\begin{split}
  T_F(z)\, \RsL{\pm}{P}(w,\bar{w}) = 
  \frac{P\, e^{\mp i\, \frac{\pi}{4}}}{\sqrt{2}\,(z-w)^{\frac{3}{2}}}\, 
  \RsL{\mp}{P}(w,\bar{w}) \,,\\ 
  \overline{T}_F(z)\, \RsL{\pm}{P}(w,\bar{w}) = -
  \frac{P\, e^{\pm i\, \frac{\pi}{4}}}{\sqrt{2}\, (\zb-\bar{w})^{\frac{3}{2}}}
  \, \RsL{\mp}{P}(w,\bar{w}) \,.
\end{split} 
\end{equation}
Let us therefore first consider the following states in the R-R sector of the spacelike super Liouville theory, 
\begin{equation}
    \ket{+}_s = \lim_{z,\zb \to 0}\, 
    e^{i\frac{\pi}{4}} \, 
    \RsL{+}{P}(z,\zb)\,\ket{0} \,,
    \qquad 
    \ket{-}_s =\lim_{z,\zb \to 0}\, \RsL{-}{P}(z,\zb)\,\ket{0}\,.
\end{equation}
These states satisfy
\begin{equation}
\begin{aligned}
\frac{1}{\sqrt{2}}\, (G_0^{\mathrm{sL}} + i\, \widetilde{G}^{\mathrm{sL}}_0) 
\ket{+}_s 
&= P \,\ket{-}_s\,, 
& \qquad 
\frac{1}{\sqrt{2}}\, (G^{\mathrm{sL}}_0 - i\, \widetilde{G}^{\mathrm{sL}}_0) \ket{+}_s 
&= 0\,, \\ 
\frac{1}{\sqrt{2}}\, (G^{\mathrm{sL}}_0 - i\, \widetilde{G}^{\mathrm{sL}}_0) \ket{-}_s 
&= P\, \ket{+}_s\,, 
& \qquad 
\frac{1}{\sqrt{2}}\, (G^{\mathrm{sL}}_0 + i\, \widetilde{G}^{\mathrm{sL}}_0)\ket{-}_s 
&= 0 \,.
\end{aligned}
\end{equation}

In the timelike case, the Ramond vertex operators $\RtL{\pm}{\Ph}$ again have definite fermion number. The OPE of the supercurrent with these vertex operators is given by   
\begin{equation}
\begin{split}
  T_F(z)\, \RtL{\pm}{\Ph}(w,\bar{w}) = 
  \frac{\Ph\, e^{\mp i\, \frac{\pi}{4}}}{\sqrt{2}\,(z-w)^{\frac{3}{2}}}\, 
  \RtL{\mp}{\Ph}(w,\bar{w}) \,,\\ 
  \overline{T}_F(z)\, \RtL{\pm}{\Ph}(w,\bar{w}) = -
  \frac{\Ph\, e^{\pm i\, \frac{\pi}{4}}}{\sqrt{2}\, (\zb-\bar{w})^{\frac{3}{2}}}
  \, \RtL{\mp}{\Ph}(w,\bar{w}) \,.
\end{split} 
\end{equation}
We now consider the states created by the Ramond vertex operators 
\begin{equation}
    \ket{+}_t = \lim_{z,\zb \to 0}\, 
    e^{i\frac{\pi}{4}} \, 
    \RtL{+}{\Ph}(z,\zb)\,\ket{0} \,,
    \qquad 
    \ket{-}_t =\lim_{z,\zb \to 0}\, \RtL{-}{\Ph}(z,\zb)\,\ket{0}\,.
\end{equation}
These states satisfy
\begin{equation}
\begin{aligned}
\frac{1}{\sqrt{2}}\, (G^{\mathrm{tL}}_0 + i\, \widetilde{G}^{\mathrm{tL}}_0) \ket{+}_t 
&= \Ph\, \ket{-}_t\,, & \qquad 
\frac{1}{\sqrt{2}}\, (G^{\mathrm{tL}}_0 - i\, \widetilde{G}^{\mathrm{tL}}_0) \ket{+}_t &= 0\,, \\ 
\frac{1}{\sqrt{2}}\, (G^{\mathrm{tL}}_0 - i\, \widetilde{G}^{\mathrm{tL}}_0) \ket{-}_t 
&= \Ph\, \ket{+}_t\,, & \qquad 
\frac{1}{\sqrt{2}}\, (G^{\mathrm{tL}}_0 + i\, \widetilde{G}^{\mathrm{tL}}_0)\ket{-}_t &= 0 \,.
\end{aligned}
\end{equation}

As in the NS sector, we want to construct states that satisfy the super Virasoro constraints. This requires firstly that total supercurrent zero mode, $\mathcal{G}_0$ $\widetilde{\mathcal{G}}_0$, has to annihilate the state on the worldsheet. 
The states are spanned by $\ket{\eta}_s\,\otimes \ket{\eta}_t$ for $\eta \in 
\{+,-\}$. Among these four states, a state that satisfies our requirement is  
\begin{equation}\label{eq:0BPsi}
\ket{\Psi} = \frac{1}{\sqrt{2}}\, \pqty{\ket{+}_s \, \ket{-}_t 
+ i\, \ket{-}_s\, \ket{+}_t}\,.
\end{equation}
This state is annihilated by $\mathcal{G}_0$ and  
$\widetilde{\mathcal{G}}_0$ provided 
$\Ph =-i\,P $. Furthermore, the state has, negative worldsheet matter fermion number, $(-1)^{F_M} = -1$.

With this preamble, we can describe the R-R vertex operators in the two GSO projected theories, mostly following the minimal superstring discussion of~\cite{Klebanov:2003wg}.
\begin{itemize}[wide,left=0pt]
\item In the Type 0A theory, we retain states with $(-1)^{F_\mathrm{ws}} =-1$. 
Since the ghosts contribute $(-1)^{F_\mathrm{gh}} =-1$ to the fermion number in the Ramond sector, we require states with $(-1)^{F_M} = 1$. But as we have seen above, there are none such that satisfy the physical state condition. Therefore, there are no RR states or vertex operators in the 0A theory. 
\item The aforementioned state $\ket{\Psi}$ defined in~\eqref{eq:0BPsi} survives the GSO projection in the Type 0B theory, where we instead demand $(-1)^{F_\mathrm{ws}} = 1$. Now the matter fermion number conspires with the ghost fermion number to select  $\ket{\Psi}$, which we have already noted, is annihilated by the zero modes of the supercurrent. Furthermore, since the zero-point energy in picture $(-\frac{1}{2}, -\frac{1}{2})$ is $\frac{5}{8}$, one requires 
$\hP + \hhP + \frac{1}{8} = \frac{5}{8}$, which is indeed satisfied for $\Ph = -i\,P$. So the physical state conditions are satisfied. The corresponding vertex operator in the $(-\frac{1}{2}, -\frac{1}{2})$ picture is 
\begin{equation}
  \mathcal{R}_P^{(-\frac{1}{2}, -\frac{1}{2})} 
  = g_s\, \mathfrak{c}\, \tilde{\mathfrak{c}}\, 
  e^{-\frac{1}{2}\,\varphi - \frac{1}{2}\, \tilde{\varphi}} \,\frac{1}{\sqrt{2}} \,\pqty{
  \RsL{+}{P} \, \RtL{-}{\Ph = -i\,P}  
  + i\,\RsL{-}{P} \, \RtL{+}{\Ph = -i\,P} }  \,.
\end{equation}
\end{itemize}

 All in all, we have two GSO projected super Virasoro minimal string theories, Type 0A and 0B $\sVMS$. They both have a common NS sector with the vertex operator $\mathcal{V}_P$, but only the 0B version has a R-R vertex operator $\mathcal{R}_P$. We shall briefly discuss the worldsheet observables built from these below. 

\subsection{The quantum string: basic observables}\label{sec:3ptamp}

The basic observables we can consider in the string worldsheet are the correlation function of the vertex operators. These, in analogy, with the VMS one expects should map to the dual matrix model observables that compute volumes of supermoduli spaces. For the present, we will simply examine the sphere three-point amplitude. Along the way, we will point out some subtleties in our identification of the 
structure constants, viz., signs that are not completely fixed by the conformal bootstrap logic. 

\paragraph{NS sector 3-point function:} The first observable we consider is the worldsheet 3-point function of NS operators, which should be common to both the 0A and 0B theory. 

We recall that in computing the genus-0 string amplitude need to fix the worldsheet conformal Killing vectors, and soak up the supermoduli. The former can be done as usual by exploiting the global $\mathrm{PSL}(2,\mathbb{C})$ to place three of our (unintegrated) vertex operators at $z= 0$, $z=1$ and $z=\infty$, respectively. However, only two of the fermionic moduli can be fixed by introducing the ghosts. Working with picture number symmetry, we need the sphere correlator to have net picture number $-2$. The way out is to insert an appropriate number of picture changing operators. We shall insert these at the location of one of the vertex operators. At the level of the 3-point function, we are therefore led to consider 
\begin{equation}
\mathscr{A}_{0,3} = N\, \expval{\mathcal{V}_{P_1}^{(-1,-1)}(0)
\, \mathcal{V}_{P_2}^{(-1,-1)}(1)\, \mathcal{V}_{P_3}^{(0,0)}(\infty)} \,.
\end{equation}
The contributions from the ghosts factor out, leaving behind the interesting piece that comes from the super Liouville theories
\begin{equation}
\begin{split}
\mathscr{A}_{0,3} 
&\propto 
\expval{
V_{P_1}(0)\, \Vh_{-i\,P_1}(0)\, 
V_{P_2}(1)\, \Vh_{-i\,P_2}(1)\, \pqty{
V_{P_3}(\infty)\, \Wh_{-i\,P_3}(\infty) +
W_{P_3}(\infty)\, \Vh_{-i\,P_3}(\infty)}} \\ 
&= 
\CV{b}(P_1,P_2,P_3)\, \CWh{\bh}(-i\,P_1, -i\,P_2,-i\,P_3) 
+ \CW{b}(P_1,P_2,P_3)\, \CVh{\bh}(-i\,P_1, -i\,P_2,-i\,P_3) \\
&= 
2i\, \etaW\, \frac{\CV{b}(P_1,P_2,P_3)}{\CV{b}(P_1,P_2,P_3)} 
+ 2i\, \frac{\CW{b}(P_1,P_2,P_3)}{\CW{b}(P_1,P_2,P_3)} \\
&= 2i\, (1+\etaW) \,.
\end{split}
\end{equation}
A few comments are in order: in the first line we used the  vertex operator in the $(0,0)$ picture~\eqref{eq:NSvo00}, but dropped the terms involving the operators $\Lambda_{P}$ or $\Lambdah_{-i\,P}$ since there are no 3-point functions with odd number of operators with non-vanishing spin. The second line writes out the answer in terms of the structure constants of the spacelike and timelike theories. Finally, we used~\eqref{eq:tLCs} where we plugged in the appropriate value of momenta and used $\bh = b$ to write out the expression in the third line solely in terms of the spacelike super Liouville structure constants. 

Happily, the momentum dependence cancels, as in the bosonic VMS. The result for $\mathcal{A}_{0,3}$, however, depends, on the undetermined parameter $\etaW$ in our expression for the timelike structure constant $\CWh{\bh}$. We can ensure the vanishing of 
this three-point amplitude by picking  
\begin{equation}
    \etaW = -1\,.
\end{equation}
The sign drops out of our crossing symmetry checks, so a priori we are making an assumption largely guided by the results of the matrix model calculation in~\cite{Stanford:2019vob}. One (weak) justification is the following: the timelike  theory was defined classically by an analytic continuation of the fields (see discussion above~\eqref{eq:tLsuperfield}), so there are additional phases in the vertex operators, which, one could argue, translates into a sign in $\CWh{\bh}$. It would, however, be desirable to fix this sign from first principles without resorting to this legerdemain.

\paragraph{Mixed NS and R sector 3-point functions:}
Let us next turn to correlators involving the Ramond vertex operators. We will denote these observables as $\mathscr{B}_{g,n}^{(m)}$, with the superscript indicating the number of Ramond insertions.

When the number of Ramond insertions is odd, the correlators vanish. So there is only one 3-point function  to consider, viz., 
\begin{equation}
\mathscr{B}^{(2)}_{0,3} 
= \expval{\mathcal{V}_{P_1}^{(-1,-1)}(0)\,
\mathcal{R}_{P_2}^{(-\frac{1}{2},-\frac{1}{2})}(1)\, 
\mathcal{R}_{P_3}^{(-\frac{1}{2},-\frac{1}{2})}(\infty)
}\,.
\end{equation}
Now the picture number is saturated to the required background value of $-2$ without need for any picture changing operator.  

Factoring out the ghost contributions again, we are left with a product of correlators in the spacelike and timelike super Liouville theories.  To wit, 
\begin{equation}
\begin{split}
\mathscr{B}^{(2)}_{0,3} 
&\propto 
\expval{
V_{P_1}(0) \Vh_{-i P_1}(0)\,
\pqty{\RsL{+}{P_2} \,\RtL{-}{-i P_2} + i\, 
\RsL{-}{P_2} \,\RtL{+}{-iP_2}}(1)\, 
\pqty{\RsL{+}{P_3}\, \RtL{-}{-iP_3} + i\, 
\RsL{-}{P_3} \,\RtL{+}{-i P_3}}(\infty)
}\\
&= 
\expval{
V_{P_1}(0)\,\RsL{+}{P_2}(1)\, \RsL{+}{P_3}(\infty)}\,
\expval{\Vh_{-i\,P_1}(0)\,\RtL{-}{-i\,P_2}(1)\, \RtL{-}{-i\,P_3}(\infty)} \\ 
&\qquad \qquad - 
\expval{
V_{P_1}(0)\,\RsL{-}{P_2}(1)\, \RsL{-}{P_3}(\infty)}\,
\expval{\Vh_{-i\,P_1}(0)\,\RtL{+}{-i\,P_2}(1)\, \RtL{+}{-i\,P_3}(\infty)} \\
&= 
\frac{1}{4} \pqty{\Ce{b}(P_1,P_2,P_3) + \Co{b}(P_1,P_2,P_3)}\, \pqty{\Ceh{\bh}(-iP_1,-iP_2,-iP_3) - \Coh{\bh}(-iP_1,-iP_2,-iP_3)}\\
& \qquad \qquad 
- \frac{1}{4} \pqty{\Ce{b}(P_1,P_2,P_3) - \Co{b}(P_1,P_2,P_3)}\, \pqty{\Ceh{\bh}(-iP_1,-iP_2,-iP_3) + \Coh{\bh}(-iP_1,-iP_2,-iP_3)} \\ 
&= 
\frac{1}{4} \pqty{\Ce{b}(P_1,P_2,P_3) + \Co{b}(P_1,P_2,P_3)}\, \pqty{\frac{1}{\Co{b}(P_1,P_2,P_3)} - \frac{\etaR}{\Ce{b}(P_1,P_2,P_3)}}\\
& \qquad \qquad 
- \frac{1}{4} \pqty{\Ce{b}(P_1,P_2,P_3) - \Co{b}(P_1,P_2,P_3)}\, \pqty{\frac{1}{\Co{b}(P_1,P_2,P_3)} + \frac{\etaR}{\Ce{b}(P_1,P_2,P_3)}} \\
&=
\frac{1}{2} (1 -\etaR )\,.
\end{split}
\end{equation}
We have once again written out the correlators of the two theories in terms of the structure constants. The identification of the timelike structure constants in~\cref{eq:tLCs} coupled with their evaluation at 
$\Ph_i = -i\, P_i$ and $\bh =b$ leads to a perfect cancellation of the momentum dependence. 

In this case, the choice 
\begin{equation}
    \etaR = 1 \,,
\end{equation}
will result in a vanishing answer for $\mathscr{B}_{0,3}^{(2)}$. This is consistent with our choice for $\etaW$ earlier, in that one expects the analytic continuation from the spacelike to the timelike theory to give the same phases to both the fermions.  
 
\paragraph{Higher point amplitudes:} Ideally, we would like to go on and compute higher point worldsheet amplitudes. In the NS sector, we expect to be able to write down a generalization $\mathscr{A}_{g,n}$ for the genus-$g$, $n$-point amplitude. For the Type 0B theory, these are expected to vanish. In the Type 0A theory, they are non-vanishing for $g>0$~\cite{Stanford:2019vob}.  

To see the vanishing of volumes, recall that the computation of an $n$-point function of NS vertex operators on a genus-$r$ Riemann surface requires fixing $2\,(3\,g-3 + n_s)$ bosonic moduli, and $2\,(2\,g-2 + n_s)$ fermionic moduli. The former are accounted for by $\mathfrak{b}$ ghost insertions, while the latter require insertions of picture changing operators (PCO). For genus-$0$ is the number of fermionic moduli exceeds the number of bosonic moduli by one. So the volume of the supermanifold vanishes. 

In the Type 0B theory, we can also compute correlators with Ramond operators. The resulting observables $\mathscr{B}^{(m)}_{g,n}$, with $m\leq n$ denoting the number of Ramond operator insertions, are also expected to vanish to all orders in perturbation theory from the analysis of~\cite{Stanford:2019vob}. Note that in this case the vanishing is not due to the presence of extra fermionic moduli, but rather attributed to the vanishing of the disk and trumpet amplitudes in the JT supergravity case. 

We are thus far unable to establish the vanishing of $\mathscr{A}_{g,n}$ or $\mathscr{B}^{(m)}_{g,n}$. As a general principle, we expect there to be a simple argument involving the picture changing operators to indicate the vanishing of these amplitudes, without the need for detailed computation of the worldsheet correlators.\footnote{We thank Atakan Firat for helpful suggestions and discussions on this point.} We note that a brute force computation a priori appears to be both challenging and perhaps, importantly, unilluminating. For example, $\mathscr{A}_{0,4}$ involves two operators in the $(-1,-1)$ picture and two in the $(0,0)$ picture. There are now potentially non-vanishing contributions from four-point functions involving two operators with non-vanishing spin. While all of these can be deduced using the elementary building blocks and superconformal Ward identities (see for example~\cite{Suchanek:2009ths}), it is not  transparent to us how the resulting combinations end up giving vanishing observables.

\section{Discussion}\label{sec:discuss}

The worldsheet description of the VMS is an interesting non-critical string background. Despite the ingredients in the construction being non-trivial interacting (non-compact) CFTs, the physical observables, viz., the string amplitudes are simple. As demonstrated in~\cite{Collier:2023cyw} the worldsheet amplitudes compute volumes of bordered Riemann surfaces. With the identification of the geodesic boundary proper lengths with the Liouville momenta, such amplitudes end up being simple polynomial functions despite the Liouville correlation functions having complicated analytic behavior. In large part, the simplification owes to the interplay between the spacelike and timelike Liouville theories, with non-trivial relations between the structure constants of the two. 

Inspired by these developments, our investigations have focused on the $\mathcal{N}=1$ generalization. While the spacelike theory with $c> \frac{27}{2}$ has been well-studied, there hasn't been much focus on the regime $c< \frac{3}{2}$, which we have referred to as the timelike regime. Our primary results include a characterization of the non-unitary SCFT in this regime of parameters. Specifically, we have obtained the spectrum and the three-point structure constants of the timelike super-Liouville theory. The essential idea in deriving these was to exploit the constraints from degenerate 4-point functions and show that there is a second solution for them, which defines the timelike theory. While we obtained our results by exploiting these constraints, we also were able to independently numerically verify that our predictions are consistent with crossing symmetry in certain sectors. The complete verification of our result using bootstrap constraints should be feasible. Our derivation of the structure constants, and their verification, turns out to be insensitive to overall signs/phases, which we argued is consequential for matrix/string dualities. 

We also undertook a preliminary analysis of the worldsheet super Virasoro minimal string, $\sVMS$, which is constructed by coupling together the timelike and spacelike Liouville SCFTs to worldsheet gravity and implementing a diagonal GSO projection. We outlined the spectrum for the  Type 0A and 0B theories, thus constructed. The 0A theory has only NS sector states, which it shares with the 0B theory. The 0B theory has in addition an R sector state.
We analyzed worldsheet 3-point amplitudes  involving either 3 NS sector operators or one NS sector and two R sector operators. Taking inspiration from the matrix model dual to JT supergravity, which our worldsheet theory should reduce to in the classical limit, we argued for particular choices for the unfixed signs/phases in our structure constants. In particular, assuming that the genus-0 3-point amplitudes vanish, we were able to completely characterize our structure constants. 

There are several questions that remain to be addressed.  First, it would also be helpful to clarify our choice of signs in the timelike super Liouville structure constants. A priori there appears nothing wrong with either choice of sign for $\etaW$ or $\etaR$. Our choice in constructing the worldsheet theory was predicated by the duality with the matrix model. Nevertheless, it remains to be understood what the choice implies, and what the resulting non-vanishing string amplitudes compute.\footnote{We thank Lorenz Eberhardt for raising this question and suggesting that our choice of sign may be tied to the definition of the matter supercharge on the worldsheet.}

The prediction from the matrix models dual to JT supergravity is vanishing perturbative amplitudes in the Type 0B theory. We have not yet discerned whether this holds for the worldsheet amplitudes. As we note earlier in the text, it would be remarkable to have a string construction with all its perturbative amplitudes vanishing. This calls for a simple explanation, one that we hope to furnish in the near future. Once verified, this worldsheet theory would be an interesting setting to understand non-perturbative effects. On the other hand, in Type 0A theory, one could have non-vanishing perturbative amplitudes since one computes the volume of the moduli spaces weighted by the fermion parity.  It would be good to understand how to recover such from a worldsheet perspective~\cite{Muhlmann:2025wip}.

In addition to computing the higher-point amplitudes, it would also be useful to examine the dynamics of boundary states. A preliminary analysis of disk, trumpet, and double trumpet amplitudes suggests agreement with matrix model results. However, the choice of  boundary conditions for the timelike super Liouville theory needs to be better understood, in order to justify these results fully.  Finally, it would also be intriguing to examine the interpretation of the worldsheet construction from a target space perspective as a time-dependent cosmological background along the lines proposed in~\cite{Rodriguez:2023kkl} for the bosonic case.\footnote{ We would like to thank Scott Collier, Henry Maxfield, and Victor Rodriguez for discussions on these issues.}

\section*{Acknowledgements}

It is a pleasure to thank Chi-Ming Chang, Atakan Firat,  Emil Martinec, Henry Maxfield, Sameer Murthy, Douglas Stanford,  Edward Witten, and Zhenbin Yang for productive  discussions and clarifying comments. 
Furthermore, we would like to acknowledge valuable exchanges with Scott Collier, Lorenz Eberhardt, Beatrix M\"uhlmann, 
Vladimir Narovlansky, Victor Rodriguez, Ioannis Tsiares, and Joaquin Turiaci.  We thank them for their comments on a draft of the paper and for sharing their insights. 

M.R.~is supported by U.S.~Department of Energy grant DE-SC0009999. J.Z.~would like to acknowledge the hospitality of QMAP, UC Davis during a summer research internship, which was supported by the Department of Physics, Tsinghua University, through the Tsinghua Xuetang Talents Program.

\appendix

\section{Summary of special functions}\label[appendix]{sec:splfns}

In this appendix, we collect some information about the special functions that
one encounters in the super Liouville theory. The building blocks of these are
the double-Gamma and Upsilon functions familiar from the bosonic Liouville
theory. An extremely useful resource on these functions and their properties is~\cite{Eberhardt:2023mrq}.

The double-Gamma function is characterized by the functional
relations (we assume $b \in \mathbb{R}_{>0}$)
\begin{equation}\label{eq:dGfrel}
\begin{split}
\dGb{z+b^\epsilon}
 & =
\frac{\sqrt{2\,\pi}\, b^{\epsilon\,b^\epsilon\,z - \frac{\epsilon}{2}}}{\Gamma(b^\epsilon\,z)}\,
\dGb{z} \,, \qquad \epsilon \in \{-1,1\}\,.
\end{split}
\end{equation}
The function can be given an integral representation in the half-plane ($\Re(z)
  <0$) as
\begin{equation}\label{eq:dGint}
\log\dGb{z} = \int_0^\infty\, \dfrac{dt}{t}\, \bqty{
  \frac{e^{\frac{t}{2}\,(Q-2\,z)}-1}{4\, \Prod{\epsilon=\pm1 }{}
    \sinh(\frac{1}{2}\, b^\epsilon\, t)} - \frac{(Q-2\,z)^2}{8}\, e^{-t}
  - \frac{Q-2\,z}{t}}\,.
\end{equation}

Note that the relations~\eqref{eq:dGfrel} imply a recursion for
$\dGb{z + m\,b + n\, b^{-1}}$ with $m, n\in \mathbb{Z}_{\geq 0}$.
These in turn indicate that the function has simple
poles at $z =-m\,b - n\,b^{-1}$. The residue at the poles is
\begin{equation}
\begin{split}
\underset{z=-m\,b - n\,b^{-1}}{\Res}\, \dGb{z}
 & = \, \dGb{Q}\,
\frac{b^{\frac{m\,(m+1)}{2}\,b^2 -\frac{n\,(n+1)}{2}\,b^{-2} -m\,n+ \frac{m-n}{2}}}{(2\pi)^{1+\frac{m+n}{2}}} \, \prod_{j=1}^m\, \Gamma(-b^2\,j)\,
\prod_{k=1}^n\, \Gamma(-m - k\,b^{-2})\,.
\end{split}
\end{equation}

For the super Liouville theory, we define two generalizations $\dGNS{b}{z}$ and $\dGR{b}{z}$ in~\eqref{eq:dgNSR}. The functional relations~\eqref{eq:dGfrel} imply that these new functions satisfy
\begin{equation}
\begin{split}
\frac{\dGNS{b}{z+Q}}{\dGNS{b}{z}}
 & =
2\pi\, \frac{b^{1+\frac{z}{2}\,(b-b^{-1})}}{
\Gamma\pqty{1+\frac{b\,z}{2}}\, \Gamma\pqty{\frac{b^{-1}\,z}{z}}}\,, \\
\frac{\dGR{b}{z+Q}}{\dGR{b}{z}}
 & =
2\pi\, \frac{b^{\frac{z}{2}\,(b-b^{-1})}}{
\Gamma\pqty{\frac{1}{2}+\frac{b\,z}{2}}\, \Gamma\pqty{\frac{1}{2}+\frac{b^{-1}\,z}{2}}}\,.
\end{split}
\end{equation}

The properties of the NS and R Upsilon functions introduced in the text~\eqref{eq:upsNSR} can be inferred directly from those of these modified double-Gamma functions. For example, since the usual Upsilon function defined in~\eqref{eq:upsb} satisfies the relation,
\begin{equation}\label{eq:Ubfrel}
\Ub{z+b^\epsilon}= b^{\epsilon -2\,\epsilon\,b^\epsilon\,z}\,\gamma\pqty{b^\epsilon\,z}\, \Ub{z}\,, 
\end{equation}
it follows that 
\begin{equation}\label{eq:UNSRfrel}
\frac{\UNS{b}{z+b}}{\UR{b}{z}} = b^{-b\,z}\, \gamma\pqty{\frac{b\,z+1}{2}} \,,
\qquad 
\frac{\UR{b}{z+b}}{\UNS{b}{z}} = b^{1-b\,z}\, \gamma\pqty{\frac{b\,z}{2}} \,,
\end{equation}
We also make use of the oft used convention for the ratio of gamma functions, as noted in~\eqref{eq:gammadef}. 

\section{Normalization conventions in spacelike super Liouville}\label[appendix]{sec:norms}

The structure constant we use in the NS sector $\CV{b}$ is related to the
ones originally obtained up to an overall factor, which can be absorbed into the
normalization of the operators. To be
specific, consider the results obtained in~\cite{Poghossian:1996agj,Rashkov:1996np, Fukuda:2002bv}. The NS sector structure constants as presented in~\cite{Fukuda:2002bv} are 
\begin{align}\label{eq:cvwdozz}
\CVdozz(P_1,P_2,P_3)
 & = \pqty{\prod_{i=1}^3\, \NNS(P_i)}\,
\prod_{\epsilon_{2,3} = \pm 1}\,
\frac{1}{\UNS{b}{\frac{Q}{2} +i\, P_1 +i\, \epsilon_2\,P_2 +i\, \epsilon_3\, P_3}}\,, \\
\CWdozz(P_1,P_2,P_3)
 & = 2i\,\pqty{\prod_{i=1}^3\, \NNS(P_i)}\,
\prod_{\epsilon_{2,3} = \pm
  1}\,\frac{1}{\UR{b}{\frac{Q}{2} +i\, P_1 +i\,\epsilon_2\,P_2 +i\, \epsilon_3\, P_3}}\,,
\end{align}
where
\begin{equation}\label{eq:NPdef}
\NNS(P) =
\pqty{\UNSo\,\bqty{ \pi\,\mu\, b^{1-b^2}\,
    \gamma\pqty{\frac{b\,Q}{2}} }
  ^{-\frac{Q+3\,i\,P}{2\,b} }}^{\frac{1}{3}}\, \UNS{b}{Q + 2\,i\,P}\,.
\end{equation}

One can check that
\begin{equation}
\begin{split}
\CV{b}(P_1,P_2,P_3)
 & =
N_{_\mathrm{DOZZ}}\, \CVdozz(P_1,P_2,P_3)\,
\prod_{i=1}^3\,\pqty{\sNS(P_i)\,\rhoNS(P_i)}^{-\frac{1}{2}}\,, \\
\CW{b}(P_1,P_2,P_3)
 & =
N_{_\mathrm{DOZZ}}\,  \CWdozz(P_1,P_2,P_3)\,
\prod_{i=1}^3\,\pqty{\sNS(P_i)\,\rhoNS(P_i)}^{-\frac{1}{2}}\,.
\end{split}
\end{equation}

The normalization factor $\mathsf{N}$ is momentum-independent, and can be determined to be
\begin{equation}
N_{_\mathrm{DOZZ}}  =  \frac{2\, \dGNS{b}{Q}^3}{\dGNS{b}{2\,Q}}\, \pqty{\pi\,\mu\,
  \gamma\pqty{\frac{b\,Q}{2}}}^{-\frac{Q}{2\,b}}\, \UNSo\,.
\end{equation}
Here, we invoke reflection amplitude $\sNS(P)$ and $\sR(P)$. In the conventional normalization the former was determined in~\cite{Fukuda:2002bv} to be
\begin{equation}
\begin{split}
\sNS(P)
 & =
\bqty{\pi\,\mu\, b^{1-b^2}\,
  \gamma\pqty{\frac{b\,Q}{2}}}^{-\frac{2\,i\,P}{b}}\,
\frac{\UNS{b}{Q + 2\,i\,P}}{\UNS{b}{Q-2\,i\,P}} \\
& =
- \pqty{\frac{2\pi b\, \bqty{\pi\,\mu\,\gamma\pqty{\frac{b\,Q}{2}}}^{-\frac{i\,P}{b}}}{\Gamma(-i b^{-1}\,P)\,\Gamma(1-i bP)}}^2 \,
\,\frac{1}{\rhoNS(P)}\,
\end{split}
\end{equation}
Similarly, the reflection amplitude in the Ramond sector is given as
\begin{equation}
\begin{split}
\sR(P)
 & =
\bqty{\pi\,\mu\, b^{1-b^2}\,\gamma\pqty{\frac{b\,Q}{2}}}^{-\frac{2\,i\,P}{b}}\,
\frac{\UR{b}{Q + 2\,i\,P}}{\UR{b}{Q-2\,i\,P}}  \\
 & =
\pqty{\frac{2\pi \, \bqty{\pi\,\mu\,\gamma\pqty{\frac{b\,Q}{2}}}^{-\frac{i\,P}{b}}}{\Gamma(\frac{1}{2}-i b^{-1}\,P)\,\Gamma(\frac{1}{2}-i bP)}}^2 \,
\,\frac{1}{\rhoR(P)}\,.
\end{split}
\end{equation}
We have expressed this in terms of the Ramond sector density of states introduced in~\eqref{eq:rhor}.  The conventionally normalized mixed structure constants are determined in~\cite{Fukuda:2002bv} as
\begin{equation}
\begin{split}
\Cedozz
 & = \frac{\NNS(P_1)\, \NR(P_2)\, \NR(P_3)}{\Prod{\epsilon_2,\epsilon_3=\pm1}{}
  \, \UR{b}{\frac{Q}{2} +i\,P_1 +i\,\epsilon_2\,(P_2+P_3)}\, \UNS{b}{\frac{Q}{2} +
i\,P_1 +i\,\epsilon_3 \,(P_2-P_3)}} \,,                                         \\
\Codozz
 & = \frac{\NNS(P_1)\, \NR(P_2)\, \NR(P_3)}{\Prod{\epsilon_2,\epsilon_3=\pm1}{}
  \, \UNS{b}{\frac{Q}{2} +i\,P_1 +i\,\epsilon_2\,(P_2+P_3)}\, \UR{b}{\frac{Q}{2} +
    i\,P_1 +i\,\epsilon_3\,(P_2-P_3)}} \,.
\end{split}
\end{equation}
Here $\NR(P)$ is defined analogously to~\eqref{eq:NPdef} with $\UNS{b}{z}
  \to \UR{b}{z}$. One may again verify that
\begin{equation}
\begin{split}
\Ce{b}(P_1,P_2,P_3)
 & =
N_{_\mathrm{DOZZ}} \, \Cedozz(P_1,P_2,P_3)\,
\pqty{\sNS(P_i)\,\rhoNS(P_i)\, \prod_{i=2}^3\,\sR(P_i)\,\rhoR(P_i)\, }^{-\frac{1}{2}}\,, \\
\Co{b}(P_1,P_2,P_3)
 & =
N_{_\mathrm{DOZZ}} \, \Codozz(P_1,P_2,P_3)\,
\,\pqty{\sNS(P_i)\,\rhoNS(P_i)\, \prod_{i=2}^3\,\sR(P_i)\,\rhoR(P_i)\, }^{-\frac{1}{2}}\,.
\end{split}
\end{equation}
%

\section{Superconformal Ward identities}\label[appendix]{sec:sward}

We collect here some useful results regarding the superconformal Ward identities, both in the NS and R sectors. 

The two and three point functions of a superconformal primary scalar superfield $\SF{S}_P(z,\theta)$ can be given compact expressions in superspace.
Given $Z_i \equiv (z_i,\zb_i,\theta_i,\thetab_i)$, we define the holomorphic combinations
\begin{equation}
\begin{split}
z_{ij}
&= 
    z_i -z_j \,, \\
\mathfrak{z}_{ij}
& = 
    z_{ij} - \theta_i\, \theta_j \,,  \\
\vartheta 
& = 
    \frac{1}{\sqrt{z_{12}\,z_{13}\, z_{23}}}\,
     \pqty{\theta_1\, z_{23} + \theta_2 \, z_{31}
    + \theta_3 \, z_{12} - \frac{1}{2}\, \theta_1\, \theta_2\,\theta_3}\,.
\end{split}
\end{equation}
The anti-holomorphic combinations are analogously defined.
Then, the superfield 2-point correlators are (with conventional normalization)
\begin{equation}
\expval{\SF{S}_{P_1}(Z_1)\, \SF{S}_{P_2}(Z_2)}
= \frac{\delta_{P_1,P_2}}{\abs{\mathfrak{z}_{12}}^{4\,h_1}} \,.
\end{equation}
The 3-point function similarly can be shown to be
\begin{equation}
\expval{\SF{S}_{P_1}(Z_1)\, \SF{S}_{P_2}(Z_2)\, \SF{S}_{P_3}(Z_3)}
=
\frac{\CV{b}(P_1,P_2,P_3) + \vartheta\, \overline{\vartheta}\,
  \CW{b}(P_1,P_2,P_3)} {\abs{\mathfrak{z}_{12}}^{2\,(h_1+h_2-h_3)}\, \abs{\mathfrak{z}_{13}}^{2\,(h_1 + h_3-  h_2)}\,
  \abs{\mathfrak{z}_{23}}^{2\,(h_2 + h_3-h_1)}}\,.
\end{equation}
Expanding out the superfield using~\eqref{eq:Snsop} we can determine
all the NS sector 3-point functions in terms of the structure constants $\CV{b}$ and
$\CW{b}$.

For correlators involving Ramond vertex operators, one has
\begin{equation}
\begin{aligned}
  \expval{\SF{S}_{P_3}(Z_3)\, \RsL{\delta}{z_1,\zb_1}\,
    \RsL{\delta}{z_2,\zb_2}}
   & =
  \frac{C^{\delta_1}(P_3,P_2,P_1)
  + \abs{z_{13}\,z_{23}\,z_{12}^{-1}} \, \theta_3\,\thetab_3
  \, \widehat{C}^{\delta_1}(P_3,P_2,P_1)
  }{\abs{z_{12}}^{2\,(h_1+h_2-h_3)}\,\abs{z_{23}}^{2\,(h_2+h_3-h_1)}\,
  \abs{z_{13}}^{2\,(h_1+h_3-h_2)}} \,, \\
  \expval{\SF{S}_{P_3}(Z_3)\, \RsL{\delta}{z_1,\zb_1}\,
    \RsL{-\delta}{z_2,\zb_2}}
   & =
  \frac{(z_{13}\,z_{23})^\frac{1}{2}\,z_{12}^{-\frac{1}{2}}\,
  \theta_3\,
  D^{\delta}(P_3,P_2,P_1) \,
  + \pqty{\zb_{13}\,\zb_{23}}^{\frac{1}{2}}\,\zb_{12}^{-\frac{1}{2}}\, \thetab_3
  \, \overline{D}^{\delta}(P_3,P_2,P_1) \,
  }{\abs{z_{12}}^{2\,(h_1+h_2-h_3)}\,\abs{z_{23}}^{2\,(h_2+h_3-h_1)}\,
  \abs{z_{13}}^{2\,(h_1+h_3-h_2)}} \,.
\end{aligned}
\end{equation}
We have
\begin{equation}
\begin{split}
C^+(P_1,P_2,P_3)
 & =
\frac{1}{2} \pqty{\Ce{b}(P_1,P_2,P_3)
+ \Co{b}(P_1,P_2,P_3)} \\
C^-(P_1,P_2,P_3)
 & =
\frac{1}{2} \pqty{\Ce{b}(P_1,P_2,P_3)
  - \Co{b}(P_1,P_2,P_3)} \,,
\end{split}
\end{equation}
as introduced in the main text. In addition, the rest of the coefficients are
related to these by superconformal Ward identities
\begin{equation}
\begin{split}
\widehat{C}^\delta(P_1,P_2,P_3)
 & =
-i\, \frac{\delta}{2} \bqty{(P_2^2 + P_3^2)\, C^\delta(P_1,P_2,P_3)
- 2\, P_2\,P_3\, C^{-\delta}(P_1,P_2,P_3) } \,,     \\
D^\delta(P_1,P_2,P_3)
 & = \frac{\,e^{-\frac{i\pi}{4}\,\delta}}{\sqrt{2}}
\bqty{P_2\, C^\delta(P_1,P_2,P_3) - P_3\, C^{-\delta}(P_1,P_2,P_3)} \,.
\end{split}
\end{equation}

Thus, all the 3-point functions are determined by the specification of $\CV{b}$,
$\CW{b}$, $\Ce{b}$, and $\Co{b}$.

\section{Recursion for superconformal blocks}\label[appendix]{sec:superrecurion}

The numerical check of our structure constants using  the crossing symmetry relations involves knowledge of the superconformal blocks for the 4-point sphere correlation function. These have been derived in a sequence of papers~\cite{Hadasz:2006qb,Belavin:2007gz,Hadasz:2007nt,Hadasz:2008dt,Suchanek:2010kq} and are well summarized in the thesis~\cite{Suchanek:2009ths}. In the spacelike case, this data was used to numerically check crossing early on  in~\cite{Suchanek:2010kq}. More recently, the NS sector data was rechecked numerically by the authors of~\cite{Balthazar:2022apu} who used this to verify the duality between the Type 0B non-critical string and the $c=1$ matrix model.  We will for the most part summarize the essential formulae, and note that our numerical implementation benefitted from the code developed in~\cite{Balthazar:2022apu} (a  summary of the NS sector blocks can be found in this reference).

The object of interest is the four point function of Virasoro primaries. This includes the superconformal primary $V_P$ or its lowest superdescendants packaged in $\SF{S}$ in the NS sector, and the Ramond primaries $\RsL{\pm}{P}$ in the Ramond sector.\footnote{ We will use the spacelike super Liouville theory for illustrative purposes in this section, noting that all the statements are generic to any $\mathcal{N}=1$ SCFT.} For any such Virasoro primary, denoted $X_i$, the 4-point function decomposes as
\begin{equation}
\begin{split}
 & \expval{X_4(z_4,\zb_4)\, X_3(z_3,\zb_3)\, X_2(z_2,\zb_2)\, X_1(z_1,\zb_1)} \\
 & \qquad  =
\frac{\Sum{h}{}\, C_{12h}\, C_{34h}\,
  \abs{\mathcal{F}(\{h_i\};h |z)}^2}{\abs{z_{24}}^{4\,h_2}\,
  \abs{z_{14}}^{2\,(h_1+h_4-h_2-h_3)}\,
  \abs{z_{34}}^{2\,(h_3+h_4-h_1-h_2)}\,
  \abs{z_{13}}^{2\,(h_1+h_3-h_2 - h_4)}}\,.
\end{split}
\end{equation}
Here $\mathcal{F}(\{h_i\};h|z) \equiv \mathcal{F}(h_4,h_3,h_2,h_1 ; h |z)$
is the conformal block, and $z$ is the cross-ratio
\begin{equation}
z = \frac{z_{12}\,z_{34}}{z_{13}\, z_{24}}\,.
\end{equation}
We seek the expansion of $\mathcal{F}$ in powers of the cross-ratio, given the external operators (which are specified by the weights $h_i$).

The expansion in terms of the cross-ratio, we recall, only converges within the unit disc $\abs{z} <1$. However, the sphere 4-point function has an infinite radius of convergence in two dimensions. This can be made explicit, by extending the domain of analyticity by mapping the four-punctured sphere $T^2/\mathbb{Z}_2$, cf,~\cite{Zamolodchikov:1987avt, Maldacena:2015iua}. The latter is parameterized by the elliptic nome $q$,\footnote{ The inverse map is given in terms of theta functions $z = \frac{\vartheta_2^4(q)}{\vartheta_3^4(q)}$.}
\begin{equation}
q = \exp(-\pi\,\frac{K(1-z)}{K(z)}) \,, \qquad
K(z) = {}_2F_1\pqty{\frac{1}{2},
  \frac{1}{2},1, z} \,.
\end{equation}
We present the results for the superconformal block $\mathcal{F}(\{h_i\}; h|z)$ first, and later indicate the translation to the so-called elliptic block $H(\{h_i\} ; h |q)$.

For superconformal blocks, we need to make the following distinctions:
\begin{itemize}[wide,left=0pt]
  \item Specify whether the external operators are from the NS or R sectors, and delineate whether the operators are superconformal primaries or Virasoro primaries which are superdescendants.
  \item Ascertain whether the intermediate states arise from descendants at integer or half-integer level.
  \item For the Ramond sector, decide whether the factorization is into
    NS or R intermediate states. We will focus on the case where the
    intermediate states are in the NS sector.
\end{itemize}

To present the relevant formulae, we adopt the notation similar to the one used  in~\cite{Suchanek:2009ths}. Let the external operators be parameterized by purely imaginary Liouville momentum $P = i\,\sqrt{2}\, \beta_i$ so that the weights are
\begin{equation}
h_i =
\begin{dcases}
\frac{Q^2}{8} - \beta_i^2 \                 & \text{ for NS superprimary}\,, \\
\frac{Q^2}{8} + \frac{1}{16} - \beta_i^2 \  & \text{ for R primaries}\,.
\end{dcases}\end{equation}
For NS sector Virasoro primaries obtained as superdescendants we use the
symbolic notation $*h_i$ to denote the weight. For formulae that are valid
independently of the string of the external operator weights, we will use the notation  $\mathfrak{h}_i $ to indicate this universality. In other words, $\mathfrak{h}_i$ could either be $h_i$ or $*h_i$ in the NS sector. In the R sector, it will stand in for the weight and a choice of fermion number, with the latter denoted by a sign. We will give explicit expressions when it becomes necessary to disambiguate the external operators in formulae.

\subsection{NS sector recursion in central charge}\label[appendix]{sec:NScrecurse}

In the NS sector there are 8 independent superconformal blocks. This is determined by whether the intermediate state is an integer or half-integer level descendant, the two cases labeled by $\mathrm{e}/\mathrm{o}$, respectively.
For either choice, we also have to specify the nature of the external operators, which could either be parameterized by $h_i$ or $*h_i$ for $i=2,3$.

We first note that expansion of the blocks in the cross-ratio has a universal expression
\begin{equation}
\begin{split}
\mathcal{F}^\mathrm{e}(\{\mathfrak{h}_i\}; h |z)
 & =
z^{h - \mathfrak{h}_1  - \mathfrak{h}_2} \,\pqty{
  1+ \Sum{m\in \mathbb{Z}_+}{}\, z^m\,
F_m(\{\mathfrak{h}_i\}; h ; c)}\,, \\
\mathcal{F}^\mathrm{o}(\{\mathfrak{h}_i\}; h |z)
 & =
z^{h - \mathfrak{h}_1  - \mathfrak{h}_2} \,
\Sum{m\in \mathbb{Z}_+- \frac{1}{2}}{}\, z^m\,
F_m(\{\mathfrak{h}_i\}; h ; c)\,,
\end{split}
\end{equation}
where, as before, $\{\mathfrak{h}_i\}$ is a shorthand for the ordered string $ \mathfrak{h}_4, \mathfrak{h}_3, \mathfrak{h}_2,
  \mathfrak{h}_1$.

The coefficients in the $z$-expansion obey a set of recursion relations, which
is what we are after. In the NS sector, we can define the
recursion in terms of the poles and residues of the superconformal block as a
function of the central charge. The coefficients in the $z$ expansion satisfy
the following recursion,
\begin{equation}
\begin{split}
F_0(\{\mathfrak{h}_i\}; h ; c)
 & = 1    \\
F_m(\{\mathfrak{h}_i\}; h ; c)
 & = \mathfrak{f}_m(\{\mathfrak{h}_i\}; h) +
\underset{r + s\; \in \; 2\, \mathbb{Z}_+}{
  \underset{1 \;< \;r \, s \;< \; 2\,m}{\sum_{r\geq 2} \, \sum_{s \geq 3}}}
\frac{R^{m}_{r,s}(\{\mathfrak{h}_i\}; h)}{c - c_{r,s}(h)}
\; F_{m-\frac{1}{2}\,r\,s}\pqty{\{\mathfrak{h}_i\}, h + \frac{r\,s}{2} ; c_{r,s}(h)}\,.
\end{split}
\end{equation}
We unpack this formula below by specifying the quantities appearing in order.

Firstly, $\mathfrak{f}(\{\mathfrak{h}\}; h)$ is the large $c$ limit of the
superconformal block, which is non-singular. The singularities of the block are isolated in poles, which occur at $c_{r,s}$ with residues $R^m_{r,s}$. We give expressions for  the large $c$ block and the residues below. An equivalent expression can be written by presenting the recursion in terms of poles in the  internal weight $ h = h_{r,s}$ (cf.~the Ramond sector discussion).

The poles can be specified by the value of the central charge, $c_{r,s}(h)$, for which the weight $h$ corresponds to that of the $(r,s)$ degenerate
representation. Since all the internal
states are in the NS sector, so the information is obtained
from the Kac determinant of the NS module, which gives
\begin{equation}
\begin{split}
c_{r,s}(h)
 & =
\frac{3}{2} + 3\, \pqty{b_{r,s}(h) + \frac{1}{b_{r,s}(h)}}^2 \,, \\
b_{r,s}(h)
 & =
-\frac{1}{r^2-1}\, \pqty{4\,h + r\, s-1 +
  \sqrt{16 \,h^2 + 8\, (r\, s -1)\, h + (r-s)^2} }\,.
\end{split}
\end{equation}

To give the expression for the residues, let\footnote{The factor of $\pdv{c_{r,s}}{h}$ originates from translating between the poles in the internal operator dimension to those in the central charge.}
\begin{equation}
A_{r,s}(h) = -\frac{1}{2} \,
\pqty{\pdv{c_{r,s}}{h}} \, \underset{p + q\;\in\; 2\, \mathbb{Z}}
{\underset{(p,q) \;\neq \;(0,0) || (r,s)}{\Prod{p=1-r}{r}\, \Prod{q=1-s}{s}}}\; \frac{\sqrt{2}}{p\, b_{r,s}(h) + q\, b_{r,s}^{-1}(h)} \,.
\end{equation}
Next, we introduce the `fusion polynomials', which take the form
\begin{equation}
\begin{split}
P^\mathrm{NS}_{r,s}(h_1,\mathfrak{h}_2)
 & =
\Prod{p=1-r}{r-1}\, \Prod{q=1-s}{s-1}
\pqty{\beta_1 - \beta_2 - \frac{p\, b_{r,s} +  q\, b^{-1}_{r,s}}{2\,\sqrt{2}}}
\pqty{\beta_1 + \beta_2 + \frac{p\, b_{r,s} +  q\, b^{-1}_{r,s}}{2\,\sqrt{2}}}
\end{split}
\end{equation}
where the product is over $p$ and $q$ is steps of $2$ with
$(p+q) - (r+s) \equiv 2 \mod 4$ when $\mathfrak{h}_2 = h_2$, but
$(p+q) - (r+s) \equiv 0 \mod 4$ when $\mathfrak{h}_2 = *h_2$.
The residue is given in terms of this data as
\begin{equation}
R^m_{r,s}(h_4,\mathfrak{h}_3,\mathfrak{h}_2,h_1; h)
=
\mathrm{sgn}_{rs}(\mathfrak{h}_3)\,
A_{r,s}(h)\,
\begin{dcases}
P_{r,s}^\mathrm{NS}(h_1, \mathfrak{h}_2) \,
P_{r,s}^\mathrm{NS}(h_4,\mathfrak{h}_3) \,,
 & \quad m \in \mathbb{Z}                 \\
P_{r,s}^\mathrm{NS}(h_1, *\mathfrak{h}_2) \,
P_{r,s}^\mathrm{NS}(h_4,*\mathfrak{h}_3) \,,
 & \quad m \in \mathbb{Z} -\frac{1}{2}\,.
\end{dcases}
\end{equation}
The convention is that $*\mathfrak{h}_i = *h_i$ if $\mathfrak{h}_i  = h_i$ and
$*\mathfrak{h}_i = h_i$ if $\mathfrak{h}_i = *h_i$. The sign is fixed again
depending on the nature of the external operator with
$\mathrm{sgn}_{rs}(h_i) = 1$ and $\mathrm{sgn}_{rs}(*h_i)= (-1)^{rs}$.

To finish specifying the recursion, we need to prescribe the large-$c$ blocks.
For $m\in \mathbb{Z}_+$ these are determined to be
\begin{equation}
\mathfrak{f}_m(h_4,\mathfrak{h}_3,\mathfrak{h}_2,h_1; h)
\begin{dcases}
\frac{1}{m!}
\frac{(h+ \mathfrak{h}_3-h_4)_m\, (h+\mathfrak{h}_2-h_1)_m}{(2\,h)_m}\,,
 & \quad m\in \mathbb{Z}_+                 \\
\frac{\mathrm{sgn}(\mathfrak{h}_3)}{\pqty{m-\frac{1}{2}}!}\,
\frac{(h+ \mathfrak{h}_3-h_4)_{m-\frac{1}{2}}\,
  (h+\mathfrak{h}_2-h_1)_{m-\frac{1}{2}}}{(2\,h)_{m+\frac{1}{2}}}\,,
 & \quad m \in \mathbb{Z} -\frac{1}{2} \,.
\end{dcases}
\end{equation}
Since $\mathfrak{h}_i \in \{h_i, *h_i  = h_i+\frac{1}{2}\}$
for $i= 2,3$ there are four
possibilities for each case. We let $\mathrm{sgn}(\mathfrak{h}_3) = +1$ when
$\mathfrak{h}_3 = h_3$ and $\mathrm{sgn}(\mathfrak{h}_3) = h_3 + \frac{1}{2}$.
Explicit expression for each of the four cases are given in~\cite{Balthazar:2022apu}.

Finally, the elliptic blocks are defined as
\begin{equation}\label{eq:ellipticH}
\begin{split}
\mathcal{F}^{\mathrm{e/o}}(h_4,\mathfrak{h}_3, \mathfrak{h}_2, h_1 ; h |z)
 & =
(16\,q)^{h-\frac{Q^2}{8}}\, z^{\frac{Q^2}{8}-h_1 - \mathfrak{h}_2}\,
(1-z)^{\frac{Q^2}{8}- \mathfrak{h}_2 - \mathfrak{h}_3} \, \\
 & \qquad\times
\vartheta_3(q)^{\frac{3\,Q^2}{2}-4\, (h_1 + \mathfrak{h}_2 + \mathfrak{h}_3 +
h_4)}\,\times \mathcal{H}^{\mathrm{e/o}}(h_4,\mathfrak{h}_3, \mathfrak{h}_2, h_1 ; h |q)\,.
\end{split}
\end{equation}
One can recast the recursion relations directly for this elliptic block. The
relevant formulae are compiled in~\cite{Suchanek:2009ths}. We followed the
strategy of computing the recursion in the $z$ variable and then translating to
the elliptic nome to extract the elliptic block.

\subsection{Ramond sector recursion in conformal weight}\label[appendix]{sec:Rhrecurse}

In the Ramond sector, the correlators can comprise either all four operators drawn from the Ramond sector, or two drawn from Ramond sector and two from the NS sector. The blocks were analyzed in~\cite{Suchanek:2010kq} for the case where all the Ramond operators had positive fermion number $\RsL{+}{P_i}$ and the NS operators were superconformal primaries $V_P$.

For the 4-point function of $\expval{\RsL{+}{P_4}(\infty)\, \RsL{+}{P_3}(1)\, \RsL{+}{P_2}(z,\zb), \RsL{+}{P_1}(0)}$ one has 8 superconformal blocks.
The distinction lies again in the level of the intermediate NS sector states, which we label by $\mathrm{e/o}$ as before. For each such choice, there are four blocks indicated by the sign of $\beta_3$ and $\beta_2$ (this is a notational contrivance, and not a priori related to the sign of these momentum labels). These blocks have a cross-ratio expansion
\begin{equation}
\begin{split}
\mathcal{F}^\mathrm{e}(\{\mathfrak{h}_i\}; h |z)
 & =
z^{h - h_1  - h_2} \,\pqty{
  1+ \Sum{m\in \mathbb{Z}_+}{}\, z^m\,
F_m(\{\mathfrak{h}_i\}; h ; c)}\,, \\
\mathcal{F}^\mathrm{o}(\{\mathfrak{h}_i\}; h |z)
 & =
z^{h - h_1  - h_2+\frac{1}{2}}
\Sum{m\in \mathbb{Z}_+- \frac{1}{2}}{}\, z^m\,
F_m(\{\mathfrak{h}_i\}; h ; c)\,,
\end{split}
\end{equation}
We now view $\{\mathfrak{h}_i\}$ as the ordered string
$\beta_4, \pm\beta_3, \pm\beta_2,\beta_1$. We focus on these correlators for
definiteness, but will note that there are 12 more blocks arising from
$\expval{V_{P_4}(\infty)\, \RsL{+}{P_3}(1)\, V_{P_2}(z,\zb)\, \RsL{+}{P_1}(0)}$
and $\expval{V_{P_4}(\infty)\, V_{P_3}(1)\, \RsL{+}{P_2}(z,\zb)\,
    \RsL{+}{P_1}(0)}$. Details of these blocks can be found
in~\cite{Suchanek:2010kq}.

Focus on correlators that factorize on NS sector intermediate state. The
coefficients $F_m$ are better expanded in the internal operator weight
\begin{equation}
F_m(\{\mathfrak{h}_i\}; h ; c)
= \mathfrak{g}_m(\{\mathfrak{h}_i\}; h)
+ \underset{r + s\; \in \; 2\, \mathbb{Z}_+}{
  \underset{1 \;< \;r \, s \;< \; 2\,m}{\sum_{r\geq 2} \, \sum_{s \geq 3}}}
\frac{R^{m}_{r,s}(\{\mathfrak{h}_i\}; h)}{h- h_{r,s}(c)}
\; F_{m-\frac{1}{2}\,r\,s}\pqty{\{\mathfrak{h}_i\}, h + \frac{r\,s}{2} ; h_{r,s}(c)}\,.
\end{equation}
While one can express the block in terms of the poles in $c$, the expansion in
$h$ is singled out by the classical limit. The issue is that the Ramond 3-point
blocks involve the rescale momenta $\beta_i$ and not the conformal weights, and
therefore depend on both $h_i$ and $c$. So the terms regular in $c$ are not
captured by the classical block.

However, there is a useful simplification: the large $h$ behavior of the blocks
can be determined. Furthermore, using it one can motivate an expansion in the elliptic variable $q$. We will skip the intermediate steps and jump directly to the latter description, referring the reader to~\cite{Suchanek:2009ths} for details.

The map to the elliptic blocks is as in~\eqref{eq:ellipticH} with all
the external weights parameterized by $\beta_i$ and $\mathfrak{h}_{2,3}$ taking
on one of two values each (labeled as $\pm \beta_{2,3}$). The statement of import
is
\begin{equation}
\begin{split}
\mathcal{H}^{\mathrm{e/o}}(\{\mathfrak{h}_i\}; h | q)
 & =
g^{\mathrm{e/o}} + \underset{r, s \; \in 2\, \mathbb{Z}}{\Sum{r,s\; >\;0}{}}\,
(16\,q)^{\frac{1}{2}\,r\,s} \, \frac{R_{r,s}(\{\mathfrak{h}_i\}; h)}{h - h_{r,s}}\,
\mathcal{H}^\mathrm{e/o}(\{\mathfrak{h}_i\}; h_{r,s} + \frac{1}{2}\, r\,s | q)
\\
 & \qquad
-  \underset{r, s \; \in 2\, \mathbb{Z}+1}{\Sum{r,s\; >\;0}{}}\,
(16\,q)^{\frac{1}{2}\,r\,s} \, \frac{R_{r,s}(\{\mathfrak{h}_i\}; h)}{h - h_{r,s}}\,
\mathcal{H}^\mathrm{o/e}(\{\mathfrak{h}_i\}; h_{r,s} + \frac{1}{2}\, r\,s |
q)\,.
\end{split}
\end{equation}
Here
\begin{equation}
g^\mathrm{e} = 1\,, \qquad g^\mathrm{o} = 0\,.
\end{equation}
The residues are in turn given as combinations of the coefficients $\hat{A}_{r,s}$ and
fusion polynomials.
\begin{equation}
R_{r,s}(\beta_4, \pm\beta_3, \pm \beta_2 , \beta_1; h)
= \hat{A}_{r,s}(c) \, P^\mathrm{R}_{r,s}(\beta_1,\pm \beta_2) \,
P^\mathrm{R}_{r,s}(\beta_4,\pm \beta_3)
\end{equation}
The coefficients $\hat{A}_{r,s}(c)$  for the recursion in weights are
\begin{equation}
\hat{A}_{r,s}(c) = 2^{r\,s-2} \,  \underset{p + q\;\in\; 2\, \mathbb{Z}}
{\underset{(p,q) \;\neq \;(0,0) || (r,s)}{\Prod{p=1-r}{r}\, \Prod{q=1-s}{s}}}\; \frac{1}{p\, b + q\, b^{-1}} \,.
\end{equation}
Finally, the fusion polynomials for the case of interest are
\begin{equation}
P^\mathrm{R}_{r,s}(\beta_1,\pm \beta_2)
=
\Prod{p=1-r}{r-1}\, \Prod{q=1-s}{s-1}
\pqty{\beta_1 \mp \beta_2 - \frac{p\, b_{r,s} +  q\, b^{-1}_{r,s}}{2\,\sqrt{2}}}
\Prod{p'=1-r}{r-1}\, \Prod{q'=1-s}{s-1}
\pqty{\beta_1 \pm \beta_2 - \frac{p'\, b_{r,s} +  q'\, b^{-1}_{r,s}}{2\,\sqrt{2}}}\,.
\end{equation}
The product again is in steps of 2 for each of the 4 variables $\{p,q,p',q'\}$.
The difference is that $(p+q) - (r+s) \equiv 2 \mod 4$ while
$(p'+q') - (r+s) \equiv 0 \mod 4$. Notice that in either case, only the rescaled momenta of the associated external operator appear in the formulae.


\providecommand{\href}[2]{#2}\begingroup\raggedright\endgroup

\end{document}